\documentstyle[preprint,aps,tighten,epsf]{revtex}

\begin{document}

\draft

\newcommand{\beq}{\begin{equation}}
\newcommand{\eeq}{  \end{equation}}
\newcommand{\bea}{\begin{eqnarray}}
\newcommand{\eea}{  \end{eqnarray}}
\newcommand{\bit}{\begin{itemize}}
\newcommand{\eit}{  \end{itemize}}

\title{Geometrical approach to the distribution of 
the zeroes for the Husimi function.
\thanks{Submitted to Journal of Physics A.}}

\author{Fabricio Toscano and 
        Alfredo M.Ozorio de Almeida}

\address{Centro Brasileiro de Pesquisas F\'{\i}sicas, \\
	 Rua Xavier Sigaud 150, CEP 22290-180,RJ,Rio de Janeiro, 
	 Brazil \\
         {\rm e-mail: toscano@cbpf.br}}

\date{\today}

\maketitle

\begin{abstract}

We construct a semiclassical expression for the Husimi function 
of autonomous systems in one degree of freedom, 
by smoothing with a Gaussian function an expression that captures 
the essential features of the Wigner function in the semiclassical 
limit. 
Our approximation reveals the ``{\it center and chord}'' 
structure that the Husimi function inherits from the Wigner
function, down to very shallow {\it valleys}, where lie the
Husimi zeroes.
This explanation for the distribution of zeroes along curves 
relies on the geometry of the classical torus, rather than the
complex analytic properties of the WKB method in the Bargmann
representation.
We evaluate the zeroes for several examples. 

\end{abstract}

\pacs{ }


\narrowtext

%
%
\section{Introduction}

The features that distinguish integrable from chaotic motion in classical
mechanics manifest themselves most clearly in phase space.
This is one of the reasons for the great interest in the so called 
``{\it quasiprobability distribution functions}'' in phase space within
the semiclassical theory of quantum states. 
These distributions are defined as the symbols associated with the 
density operator $\hat{\rho}$ in some representation of quantum operators
\cite{baljen,wigner}.  
It is expected that these representations of quantum states
show  the differences between an integrable, or a chaotic classical  
counterpart in the semiclassical limit ($\hbar \rightarrow 0$).
Among these phase space representations of quantum states, the Wigner 
function ({\it i.e.} the symbol of the density operator $\hat{\rho}$ in 
the Weyl representation) and its smoothing by a Gaussian function, 
the Husimi function, are of paramount importance.
In fact, Berry \cite{berry1} showed that the peak of the amplitude of the 
Wigner function is located very close to the curve of constant energy,
for a pure state of an autonomous system with one degree of freedom,
collapsing onto a zero-width distribution ({\it i.e.}, a delta function) 
over that curve  in the classical limit ($\hbar=0$) (see also
\cite{berrybalaz,balaz,baljen}). 
Ozorio de Almeida and Hannay generalized this picture for states 
supported by invariant tori of higher dimensions \cite{ozha}. 
In all such systems, the semiclassical analysis of the Wigner function 
\cite{berry1,ozha,ozo82,berrybalaz,balaz} reveals an interesting 
geometrical structure of ``{\it chords and centers}'' that determines 
the phase of the oscillations of the Wigner function as the  point 
${\bf x}=(q,p)$ is varied within the torus.  
This phase is proportional to the symplectic area -or {\it center} 
action- bounded by the torus and the {\it chord}, centered on 
${\bf x}$, joining two points of the torus. 
The links of this ``{\it semiclassical geometry}'' to the generating
function formalism of classical mechanics and the path integrals
of quantum mechanics are reviewed in reference \cite{alfredo}.

Although the oscillations of the Wigner function thus reflect legitimate 
structures of classical mechanics, its positive definite Gaussian smoothing,
the Husimi function, is much closer to a classical Liouville density.
The density peaks near the region of classical motion, decaying 
exponentially in classically forbidden regions
\cite{taka,kls}. 
The first impression is that smoothing cancels all trace of the 
{\it centre} and {\it chord} skeleton of the Wigner function. 
However, we shall show that very delicate effects are still discernible.  
 
First, we must recall that the Husimi function can also be viewed as the 
mean value of the density operator $\hat{\rho}$ in coherent states, 
whose holomorphic (entire) part is called the Bargmann function 
\cite{voros1}. 
This function corresponds to the wave function for the quantum state in a
representation of the quantum mechanics introduced by Bargmann \cite{barg} 
(in the case of the Heisenberg-Weyl group), where the basis
for the Hilbert space is made of coherent states not normalized to unity.
Thus, in the Bargmann representation the wave functions are holomorphic
(entire) functions of the variable 
$z=\frac{1}{\sqrt{2}}(\beta q-ip/\beta)$,
acting as a phase space coordinate.  
The analiticity of the Bargmann function compels its zeroes and those  
of the Husimi function to be isolated for 1-D systems.
Leboeuf and Voros \cite{lebvor} have shown that in many cases the 
distribution of Husimi zeroes is completely different for chaotic
maps, where they are spread out, as opposed to integrable maps, where
they are distributed along curves. 
Only the latter alternative is available for systems with continous
time and one degree of freedom.
Furthermore, these lines of zeroes cannot occur close to the 
energy shell where the smoothed Wigner function has a non-oscillatory 
peak. The lines supporting zeroes may only be found in regions
where the Husimi function is already exponentially small.

In these circunstances, we can only expect to predict the general pattern
of zeroes with a very delicate ``{\it subdominant}'' semiclassical 
theory. 
This is the case of WKB type of theory developed by Voros for the Bargmann
representation \cite{voros1}, which predicts zeroes on the anti-stokes lines
where two or more branches of the complex action have the same amplitude.
The zeroes along these lines are selected by the condition that the
imaginary part of the complex action be an integer multiple of $\pi$. 
However, this approach has practical difficulties to obtain explicit 
formulae even to leading order in $\hbar$. 
First, it is generally very difficult to obtain analitically 
the branches of the classical energy curve in complex coordinates, 
in order to calculate explicitily the complex action 
({\it i.e.} the phase in the WKB wave function).
Second, an approximation valid anywhere outside the neighborhood
of the energy curve, requires the analytic continuation of the functions
that define the branches; this needs the analyticity of the Weyl symbol, 
$H_W$, for the quantum Hamiltonian. 
 
Let us thus return to the picture of the Husimi function as a smoothing 
of the Wigner function.
Since our approximation does not maintain explicitily the analytical
properties, one could not expect to establish that there exist
isolated zeroes in this way, but we can seek for shallow 
{\it valleys}, even in the region where the Husimi function is already
exponentially small, and for oscillations along their bottom as 
indications of where the zeroes may lie. 
The simplest guess is that the two dominant regions in the evaluation 
of the Husimi function are the neighborhood of the centre of the Gaussian 
and the maximum of the Wigner function along the energy curve $\cal E$.  
Since we know that a zero will only be found when the Gaussian 
is far removed from the energy curve, we use Berry's simple 
cosine-oscillatory representation of the Wigner function
for the local approximation.
For the contribution of the region near the energy curve, we start
from the even cruder classical aproximation that the Wigner function
is a $\delta$-function along $\cal E$.
After the smoothing, the first term remains cosine-oscillatory with 
essentially the same phase (proportional to the {\it center} action, 
upon small corrections), but now damped by an exponential function 
which decreases, essentially with the length of the {\it chords} 
(upon small corrections). 
The second term is everywhere positive, smooth and peaked in the
energy curve $\cal E$. 
The combination of both produces a positive smooth expression, peaked  
along the curve $\cal E$ oscillating in a {\it valley} of 
local minima that approach the zeroes of the Husimi function when 
$\hbar \rightarrow 0$. 
This expression is valid only inside the energy curve and depends only 
on the properties of the torus $\cal E$. 

The paper is organized as follows. 
In Section \ref{SectionRWH} we summarize important results 
concerning Wigner and Husimi functions.
In Section \ref{SectionPB} we find a semiclassical expression
of the Husimi function for the case of a particle in a box,
as a simple model of the geometrical approach.   
In Section \ref{SectionGA} we introduce our geometrical 
approach to the distribution of the Husimi zeroes in 1-D systems.
In Section \ref{SectionPCF} we apply this approach to the problem
of a particle under the action of a constant force. 
This example corresponds to an unbounded problem whose convex 
energy curve is open.
Finally, in Section \ref{SectionGCQO} we present the results for the
case of a particle subject to an asymmetric anharmonic potencial, 
as example of a general system with a convex and closed energy
curve.
%


%

\section{Review of Wigner and Husimi functions}

\label{SectionRWH}


{\it Quasiprobability distribution functions} are symbols
associated with the density operator, $\hat{\rho}$, in some representation 
of quantum operators \cite{baljen,wigner}. 
The Wigner function is the Weyl symbol of the density operator. 
The symbol of an operator $\hat{\mbox{A}}$, in the Weyl representation 
is given by the function,
\beq   
\label{asymbol}
A_W({\bf x})=\int d\xi_q \;\;
\langle q+\xi_q/2 |\hat{\mbox{A}}| q-\xi_q/2 \rangle \;\;
\exp\left[ -ip\xi_p/\hbar \right]
\;
,
\eeq
where ${\bf x}=(q,p)$ (all integrations in this work run 
from $-\infty$ to $+\infty$ unless indicated). 
So, in the case of pure states in 1-D systems, the Wigner function is
\beq    
\label{defwigner}
W({\bf x})\equiv
\left(\frac{1}{2\pi\hbar}\right)
\rho_W({\bf x})=
\left(\frac{1}{2\pi\hbar}\right)
\int d\xi_q \;\;
\langle q+\xi_q/2 |\psi       \rangle
\langle \psi      | q-\xi_q/2 \rangle \;\;
\exp\left[ -ip\xi_p/\hbar \right].
\eeq
We remark that since $\mbox{Tr}[\hat{\rho}]=
\int \frac{d{\bf x}}{2\pi\hbar} \rho_W({\bf x})=1$ for normalizable states 
$|\psi \rangle$, while diverging otherwise, the prefactor in
(\ref{defwigner})  will not be considered for unbounded states.

The semiclassical analysis of this function was first developed by Berry
\cite{berry1} for the case of an eigenstate of energy $E$,  
in nonrelativistic 1-D systems, where the classical 
Hamiltonian is of the form 
\beq
\label{hamil}
\mbox{H}({\bf x})=p^2/2m + \mbox{V}(q)
\eeq
and the ``torus'' is the smooth convex curve, $\cal E$, of constant energy
($\mbox{H}({\bf x})=E$).
We briefly summarize the results in \cite{berry1}, which are
important for this work (for more details see also \cite{balaz,berrybalaz}).
 
{\bf a) {\bf \it The simple semiclassical approximation.}}

\begin{figure} 
\epsfxsize=16.0cm
\epsfbox[70 136 509 625]{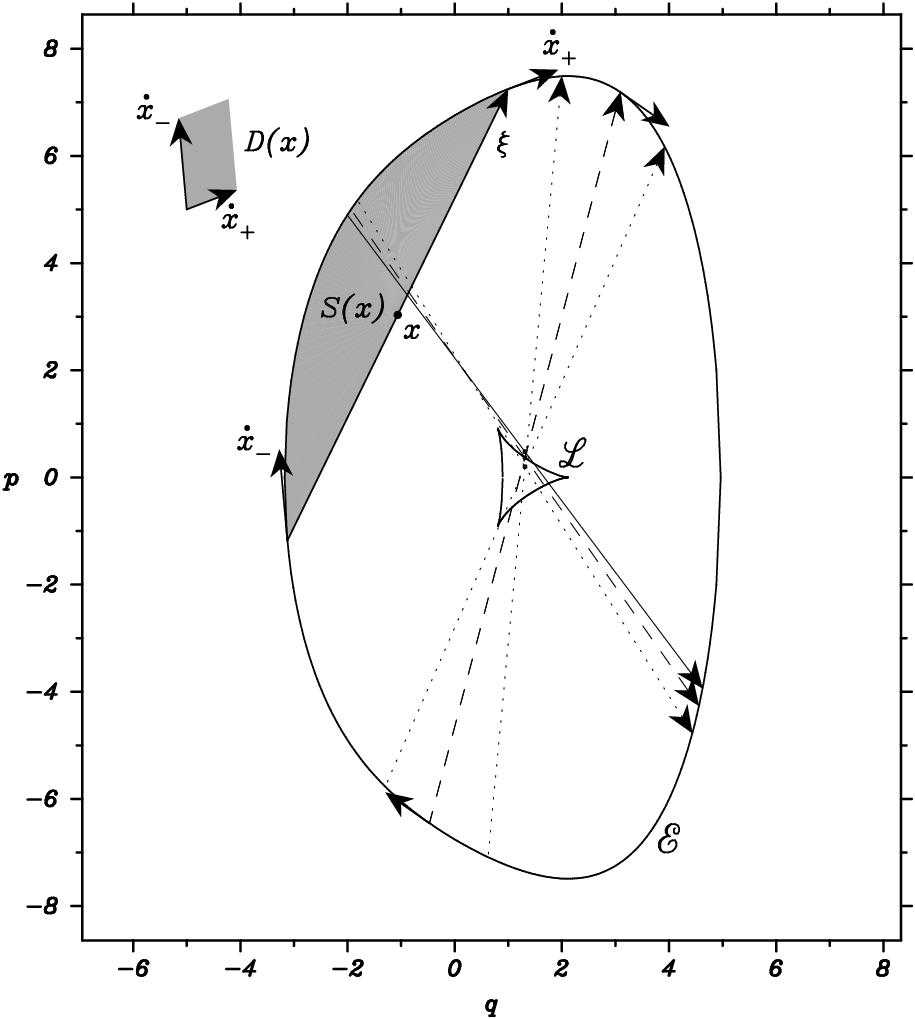}
\caption{}
\label{Fig1}
\end{figure}

This is obtained by replacing the primitive WKB functions 
({\it i.e.} the semiclassical solution of the time independent 
Schrodinger equation) in (\ref{defwigner}) and evaluating the integral 
by the stationary phase method. The result is symmetric in $q$ 
and $p$ and depends only on the geometry of the classical curve $\cal E$,
\beq   
\label{berrsimple}
W_{SCL}({\bf x})=
\frac{2}{\pi\sqrt{2\pi\hbar}(\omega^{-1})}
\sum_{\mbox{\scriptsize chord}-j}
\frac{1}{\sqrt{D_j({\bf x})}}\;
\cos\left\{ \frac{S_j({\bf x})}{\hbar}-\frac{\pi}{4} \right\}\;\;,
\eeq
where the function $S_j({\bf x})$ is the symplectic area bounded by the 
energy curve $\cal E$ and the {\it chord} ${\bf \xi}$, centered in 
${\bf x}$, that joins the points ${\bf x_+}$ and ${\bf x_-}$ on the 
``torus'' (see FIG.\ref{Fig1}).
The sum is over all the {\it chords} centered on ${\bf x}$ and
$\omega$ is the frequency of the classical motion around $\cal E$.
The $D_j({\bf x})$ are the {\it skew products} of the phase space velocities
at the tips of the {\it chord}, 
\beq   
\label{denom}
D_j({\bf x})=
{\bf \dot{x}_-}
\wedge 
{\bf \dot{x}_+} =
\dot{p}_- \dot{q}_+ - \dot{q}_- \dot{p}_+ \;\;,
\eeq
representing the area of the parallelogram formed by the pair of vectors
(the point ${\bf x_+}$ is reached after the point ${\bf x_-}$ 
in the classical motion along ${\cal E}$, see FIG.\ref{Fig1}).
Outside of the convex energy curve $\cal E$ there are no {\it chords}, so 
$W_{SCL}({\bf x})=0$.

The {\it Wigner caustic} labeled as $\cal L$ in FIG.\ref{Fig1}, is the 
border of regions with different numbers of {\it chords}: within the 
{\it caustic} there are three {\it chords}, on the {\it caustic} two 
and outside only one.  
On the {\it Wigner caustic} and on $\cal E$, generically two pairs of 
stationary points coalesce; in both cases the simple method of stationary 
phase is inapplicable.  
We stress that two of the three terms in (\ref{berrsimple}) diverge
as a smooth side of $\cal L$ is approached from the inside 
(FIG.\ref{Fig1}).
The reason is that the phase space velocities, 
${\bf \dot{x}_+}$ and ${\bf \dot{x}_-}$, are parallel, so the area 
$D_j({\bf x})$ is zero.
We also remark that when the eigenstate is normalizable, the prefactor in
(\ref{berrsimple}), that arises from  the correct normalization of the
primitive WKB function, does not give the correct normalization of the
Wigner function. 
 
{\bf b) {\bf \it The uniform approximation.}} 

Simultaneous consideration of a pairs of stationary points in
(\ref{defwigner}) yields
\beq   
\label{berruni}
W_{SCL}({\bf x})=
\frac{\sqrt{2}}{\pi\hbar^{2/3}(\omega^{-1})}
\sum_{\mbox{\scriptsize chord}-j}
\frac{1}{\sqrt{D_j({\bf x})}}\;
\left[\frac{3S_j({\bf x})}{2}\right]^{1/6}
\mbox{Ai}\left\{ -\left[\frac{3S_j({\bf x})}{2\hbar}\right]^{2/3}\right\}\; .
\eeq
This is an uniformily valid approximation not only as ${\bf x}$ moves onto 
$\cal E$, but also when ${\bf x}$ lies on the convex side of $\cal E$ 
($\mbox{H}({\bf x})> E$) where the stationary values and the function 
$S({\bf x})$ are imaginary. However, further refinements of the method 
of stationary phase are required to obtain an approximation uniformly
valid over $\cal L$.

On the concave side of $\cal E$ ($\mbox{H}({\bf x})< E$) and not too 
close to $\cal E$, $S({\bf x})$ is large in comparison with $\hbar$ 
so the Airy function can be replaced by its asymptotic form 
for large negative argument \cite{abramo},   
\beq
\label{airyapp}
\mbox{Ai}\{ -w \}\approx
\frac{1}{\sqrt{\pi}}
[w]^{-1/4}
\cos \left\{ \frac{2}{3}[w]^{3/2} - \frac{\pi}{4} \right\}
\eeq
and (\ref{berrsimple}) is recovered. On the convex side of $\cal E$ 
the Airy function has positive argument and the semiclassical Wigner 
function decays exponentially away from $\cal E$.

When the eigenstate is normalizable, the {\it uniform} Wigner function 
(\ref{berruni}) is correctly normalized.

{\bf c) {\bf \it The transitional approximation.}}

Very close to $\cal E$ an expansion of (\ref{berruni}) yields, 
\beq    
\label{berrtran}
W_{SCL}({\bf x})=
\frac{1}{2\pi(\omega^{-1})}\;
\frac{2}{[\hbar^2 B({\bf x})]^{1/3}}\;
\mbox{Ai}\left\{ 
\frac{2}{[\hbar^2 B({\bf x})]^{1/3}}\;
[\mbox{H}({\bf x})-\mbox{E}] \right\} 
\; ,
\eeq
where
\beq   
B({\bf x})=\mbox{H}_q^2 \,\mbox{H}_{pp}+\mbox{H}_p^2 \,\mbox{H}_{qq}+
2\,\mbox{H}_{pq}\,\mbox{H}_p\,\mbox{H}_q \;\; ,
\eeq
with all the partial derivatives of $\mbox{H}$ evaluated at ${\bf x}$. 
$B({\bf x})$ remains finite as ${\bf x}$ moves onto $\cal E$. 

{\bf d) {\bf \it The classical limit.}}

This corresponds to the limit when $\hbar=0$ and is obtained 
by letting $\hbar \rightarrow 0$ in (\ref{berrtran}) and using 
the result,
\beq
\label{clalim}
\lim_{\epsilon \rightarrow 0}\;
\frac{1}{\epsilon}\;
\mbox{Ai}\left(\frac{{\bf y}}{\epsilon}\right)=
\delta ({\bf y})\;\;,
\eeq
to give,
\beq    
\label{wigclalim}
W_{CL}({\bf x})=
\frac{1}{2\pi(\omega^{-1})}\;
\delta[\mbox{H}({\bf x})-\mbox{E}]\;.
\eeq
Along the {\it Wigner caustic} the modulus of the Wigner function
takes large values. 
However, the infinitely rapid oscillations along $\cal L$ cancel 
the amplitude of the delta function. 
In the case of normalizable eigenstates, 
(\ref{berrtran}) and (\ref{wigclalim}) are correctly normalized.

The prefactor in (\ref{berrsimple}) and the normalization constants 
in (\ref{berruni}), (\ref{berrtran}) and (\ref{wigclalim}), are defined
for normalizable states. 
When the states are not normalizable, the formulae are still valid,
but now it is possible to define the normalization using the 
orthogonality conditions if the wave function belongs to an orthogonal 
set \cite{balaz,baljen}.
For these cases, the classical frequency $\omega$ should not be 
included in the formulae.
An example of this type of normalization appears in Section
\ref{SectionPCF}.

From the fundamental ``{\it quasiprobability}'' property 
(see {\it e.g.} \cite{alfredo}),
\beq
\label{vmedio}
\langle \psi|\hat{\mbox{A}}|\psi\rangle=
\int d{\bf x}\; A_W({\bf x})\; W_{\psi}({\bf x})
\; ,
\eeq
we obtain that the ``scalar product'' of Wigner functions, 
\beq
\label{escprod}
\int d{\bf x}\; W_{\phi}({\bf x})\; W_{\psi}({\bf x})=
\left(
\frac{1}{2\pi\hbar}
\right)
|\langle\phi|\psi\rangle|^2
\; ,
\eeq
is always positive definite, including the projection onto positions,
\beq
\int dp W_{\psi}({\bf x})= 
|\langle q |\psi\rangle|^2
\; ,
\eeq
giving the probability density (see \cite{acota}).

Recently, Ozorio de Almeida \cite{alfredo} reviewed the link between
the ``{\it semiclassical geometry}'' underlying the Wigner function 
and the generating function formalism of classical mechanics.
This is based on canonical conjugate variables, 
the {\it centers}, ${\bf x}=(q,p)$, and the {\it chords}, 
${\bf \xi} =(\xi_q,\xi_p)$,
as an alternative in the description of the classical evolution. 
Hence, instead of specifying the 2L-dimensional initial, ${\bf x_-}$,
and final, ${\bf x_+}$, points  in phase space, we can give the vector 
${\bf \xi}$, joining each pair, and the position ${\bf x}$ of its center.
The evolution is described by the canonical transformations given by the 
{\it center} or {\it chord} generating functions ({\it i.e.} the 
{\it center} or {\it chord} actions respectively), from which we obtain the
corresponding canonical conjugate variables by differentation:
\beq
\label{derivs}
\xi_q({\bf x})=-\,\frac{\partial S({\bf x})}{\partial p}
\hspace{0.8cm}
;
\hspace{0.8cm}
\xi_p({\bf x})=\frac{\partial S({\bf x})}{\partial q}
\; .
\eeq
The {\it center} action $S({\bf x})$, for fixed energy, is the function 
in the expressions (\ref{berrsimple}) and (\ref{berruni}).


The Husimi function is another {\it quasiprobability distribution function}; 
it corresponds to the normal symbol of the density operator in the
diagonal coherent states representation \cite{voros1} (or Husimi
representation \cite{husimi}).
In this representation the normal symbol of an operator 
$\hat{\mbox{A}}$ is the expectation value,
\beq
A_N({\bf X})=
\langle \Omega_{\bf X} |\hat{\mbox{A}}|\Omega_{\bf X} \rangle \;,
\eeq  
where the $|\Omega_{\bf X} \rangle$ are the minimum uncertainty states
known as coherent states \cite{pere,klauder}.
These states are eigenstates of the {\it destruction operator},
\beq
\label{desoperator}
\hat{a}=2^{-1/2}
(\beta \hat{Q} + i \hat{P}/\beta)
\eeq
for the reference harmonic oscillator:
\beq
\label{oscref}
\hat{\mbox{H}}=\hat{P}^2/2m 
+ m\,\omega_r^2 \;\hat{Q}^2/2=
\omega_r\,\hat{a}^{\dagger}\hat{a} +\hbar/2
\;,
\eeq
with $\beta=(m\,\omega_r)^{1/2}$. They can be obtained by displacing 
the normalized ground state, 
$|\Omega_{\bf X=0} \rangle\equiv|0\rangle$, of (\ref{oscref})
to the phase space location ${\bf X}=(Q,P)$ according to,
\beq
|\Omega_{\bf X} \rangle=
\exp\{ (i/\hbar)(P\hat{Q}-Q\hat{P})\}
|0\rangle
\; .
\eeq 
Therefore, the Husimi function for pure states of  1-D system is,
\beq
\label{husdef1}
H({\bf X})=\left(\frac{1}{2\pi\hbar}\right)
|\langle \Omega_{\bf X}|\psi \rangle|^2 \;,
\eeq
where we include the prefactor, for the case of 
normalizable states, to make the Husimi function integrable to unity 
over the whole phase space.

The Husimi representation can also be viewed as a Gaussian smoothing
of the Weyl representation \cite{alfredo}, so the Husimi function
is related  to the Wigner function in the form,
\beq
\label{hugausmmoth}
H({\bf X})=\frac{1}{\pi\hbar}
\int d{\bf x} W({\bf x})
\exp\left\{ -\frac{1}{\hbar}\|({\bf x}-{\bf X})\|_{\beta}^2 \right\}
\eeq
where the ``$\beta$-metric'' is defined as 
\beq
\label{norma}
\|({\bf x}-{\bf X})\|_{\beta}\equiv [\beta^2(q-Q)^2+(p-P)^2/\beta^2]^{1/2}
\; .
\eeq
This expression is the starting point for our approach developed in the 
next sections.
Since $W_{\Omega_{\bf X}}=1/\pi\hbar 
\exp\left\{ -\frac{1}{\hbar}\|({\bf x}-{\bf X})\|_{\beta}^2 \right\} $
is the Wigner function for a coherent state (see {\it e.g.} \cite{alfredo}),
(\ref{hugausmmoth}) defines the positive definite projection, (\ref{escprod}),
of the Wigner function onto coherent states.
Note that the Weyl representation is invariant under symplectic 
transformations (linear canonical transformations). The introduction 
of a metric in (\ref{hugausmmoth}) implies that symplectic invariance
does not carry over to the Husimi function.

The fact that the coherent states are eigenstates of the {\it destruction
operator} gives them analytical properties which are translated to the 
Husimi function \cite{pere,klauder}.
We can separate the analytical part of the Husimi function if we use 
the unnormalized coherent states $|z\rangle$, such that 
$|\Omega_{\bf X} \rangle \equiv \exp\{-\bar{z}z/2\hbar\} |z\rangle$,
{\it i.e.},
\beq
\label{hubarg}
H(z)=\left(\frac{1}{2\pi\hbar}\right)
\exp\{-\bar{z}z/\hbar\}\;
|\langle z|\psi \rangle|^2 
\;
,
\eeq
where the coordinates,
\beq
z=2^{-1/2}
(\beta Q - iP/\beta)
\hspace{1cm}
,
\hspace{1cm}
\bar{z}=2^{-1/2}
(\beta Q + iP/\beta)
\;
,
\eeq
represent the complex phase space for 1-D systems. 

The function $\langle z|\psi \rangle$ corresponds to the wave function 
for the quantum state in a representation of quantum mechanics 
introduced by Bargmann \cite{barg} (in the case of the Heisenberg-Weyl 
group), where the basis for the Hilbert space are the
unnormalized coherent states (althought overcomplete).
These wave functions are holomorphic (entire) functions of the variable
$z$, which behaves as a phase space coordinate.  
From the WKB construction in the Bargmann representation, Voros \cite{voros1}
derives a semiclassical approximation for the Husimi function in 1-D 
systems. 
This is presented in Appendix \ref{CWKBM} for the energy eigenstates of 
the problem of a particle under the action of a constant force and 
the results are compared with our approximation of Section 
\ref{SectionPCF}.

%
%

\section{The particle in a box.}

\label{SectionPB}

To introduce our approach let us study the simplest
case of a particle in the symmetric classical potential:
\beq
\mbox{V}(q)= \left\{ 
\matrix{ 0  &  |q| \leq \frac{l}{2} \cr 
            &                       \cr
    +\infty &  |q| > \frac{l}{2} \cr } 
\right.
\eeq
with the pure states given by even eigenfunctions,
\beq
\label{boxeigen}
\langle q|\psi_n \rangle = 
\left\{ 
\matrix{ (2/l)^{1/2}\cos(p_n q/ \hbar) &  |q| \leq \frac{l}{2}\cr
                                       &                      \cr
                   0                   &  |q| > \frac{l}{2}   \cr }
\right.
\hspace{1cm}
\mbox{with}\;\;\;
 p_n=\pi\hbar(n+1)/l \;\;
(n\;\mbox{even})
\;\;.
\eeq
Thus the semiclassical limit for a given classical momentum, $p_n$,
corresponds to the limit of large $n$. 

The Wigner function for the state \cite{alfredo} is zero outside the
box, whereas inside:
\beq \label{wigcaja}
W(q,p)=\frac{1}{2l}
\left\{
  \frac{\sin{(2(p-p_n)y/ \hbar)}}{\pi(p-p_n)} 
+ \frac{\sin{(2(p+p_n)y/ \hbar)}}{\pi(p+p_n)} 
+ 2 \cos{(2 p_n q/ \hbar)} \frac{\sin{(2  p y/ \hbar)}}{\pi p} 
\right\}
\eeq
where $y=\frac{l}{2} - q$  if  $0\leq q \leq \frac{1}{2}$  and 
$y=\frac{l}{2}+q$  if  $-\frac{1}{2} \leq q < 0$. 

The Husimi function can be calcutated in terms of the error function 
$\Phi(z)$ (see Appendix \ref{HFPB}),
\bea
\label{hubox}
H(Q,P)&=&\frac{1}{8l\beta\sqrt{\pi\hbar}}
\left\{
e^{-\frac{(P-p_n)^2}{\hbar \beta^2}} \;
\left|\Phi(\frac{z_1}{\sqrt{2\hbar}}) + 
      \Phi(\frac{z_2}{\sqrt{2\hbar}})\right|^2 \right. 
+
e^{-\frac{(P+p_n)^2}{\hbar \beta^2}} \;
\left|\Phi(\frac{z_3}{\sqrt{2\hbar}}) + 
      \Phi(\frac{z_4}{\sqrt{2\hbar}})\right|^2  \nonumber \\
&+&
\left.
2\; e^{-\frac{(P^2+p_n^2)}{\hbar \beta^2}} \;
\Re e\left[ e^{-\frac{i2p_nQ}{\hbar}} 
       \left(\overline{\Phi(\frac{z_1}{\sqrt{2\hbar}})} + 
             \overline{\Phi(\frac{z_2}{\sqrt{2\hbar}})}\right)
       \left(\Phi(\frac{z_3}{\sqrt{2\hbar}}) + 
                       \Phi(\frac{z_4}{\sqrt{2\hbar}})\right) \right] 
\right\}
\eea 
where  the arguments of the error functions are measured from the
``corners of the phase space box'' shown in FIG.\ref{Fig2}, 
$z_1=\beta(\frac{l}{2}+Q)-i(P-p_n)/\beta$,
$z_2=\beta(\frac{l}{2}-Q)+i(P-p_n)/\beta$,
$z_3=\beta(\frac{l}{2}+Q)-i(P+p_n)/\beta$ and 
$z_4=\beta(\frac{l}{2}-Q)+i(P+p_n)/\beta$; (the overlines in (\ref{hubox})
indicate complex conjugation).
As opposed to the Wigner function, (\ref{hubox}) is not zero outside the box.  
A semiclassical analysis of this expression can be made with the help
of the asymptotic expansion of the error function $\Phi(w)$ 
for $|w|$ sufficiently large, {\it i.e.} $\hbar \rightarrow 0$
(see Appendix \ref{HFPB}). 
If we replace each $\Phi(w)$, by the first term of 
this expansion, 
\beq
\label{errofunp}
\Phi(w) \approx
\left\{
\matrix{ 1-\frac{e^{-w^2}}{\sqrt{\pi}w}& \Re e(w) \gg 1 \cr 
        -1-\frac{e^{-w^2}}{\sqrt{\pi}w}& \Re e(w) \ll -1 \cr}
\right.
\eeq
we get a good approximation of (\ref{hubox}) except in narrow 
margins along the lines $Q=-l/2$ and $Q=l/2$, that contract
when $\hbar \rightarrow 0$. 
This can be observed, for the region of interest inside the box 
and between the branches of the classical trajectory by
comparing the plots ({\bf a}) and ({\bf b}) in FIG.\ref{Fig2}.
Furthermore, the function $\Phi(w)$ approaches unity,
in the limit $|w|\rightarrow +\infty$, in the 
region $|\Im m(w)|< \Re e(w)$  ($\Re e(w) > 0$).
The intersection of these regions for each error function in (\ref{hubox})
defines a central rectangle shown in the plots {(\bf a)} and {(\bf b)} of
FIG.\ref{Fig2}. 
Hence, in the semiclassical limit the Husimi function is well represented 
by, 
\bea
\label{hurecbox}
H(Q,P)&\approx& \frac{1}{2l\beta\sqrt{\pi\hbar}}
\left\{ e^{-\frac{(P-p_n)^2}{\hbar \beta^2}} +
        e^{-\frac{(P+p_n)^2}{\hbar \beta^2}} +
    2\; e^{-\frac{(P^2+p_n^2)}{\hbar \beta^2}} \;
        \cos(2p_nQ/ \hbar)
\right\}= \nonumber \\
&=&\frac{1}{l\beta\sqrt{\pi\hbar}}
e^{-\frac{P^2}{\hbar \beta^2}}
e^{-\frac{p_n^2}{\hbar \beta^2}}
\left\{ 
\cosh(2p_nP/ \beta^2\hbar)+\cos(2p_nQ/ \hbar)
\right\}
\; ,
\eea 
within that rectangle.
This expression is explicitily positive everywhere except in its 
zeroes that lie on the axis $P=0$ where it simplifies to
\beq
\label{hupebox}
H(Q,P=0)\approx \frac{2}{l\beta\sqrt{\pi\hbar}}
e^{-\frac{p_n^2}{\hbar \beta^2}}
\left\{ \cos^2(p_nQ/ \hbar)
\right\}.
\eeq 

Alternatively, since the Husimi function is a Gaussian smearing of 
the Wigner function (\ref{hugausmmoth}), it is possible to obtain a
semiclassical approximation of (\ref{hubox}) by performing the 
Gaussian smoothing over an expression that mimics the behavior of the
Wigner function in the semiclassical limit. 
With the help of the limiting form of the delta funtion,
$\delta(x-x_0)=\lim_{L\rightarrow +\infty}\sin[L(x-x_0)]/\pi(x-x_0)$, 
we observe that, semiclassically, the skeleton of the Wigner function (\ref{wigcaja}) is 
\beq
\label{humimic}
W(q,p)\sim \frac{1}{2l}
\left\{ 
\delta (p-p_n) + \delta (p+p_n) + 2\; \delta(p) \;\cos(2p_nq/\hbar)
\right\}
\; .
\eeq
Replacing this expression in (\ref{hugausmmoth}) and limitting 
the integration to the range $(-l/2,l/2)$, leads to
\bea
\label{hupboxn}
H_{SCL}(Q,P)&=&\frac{1}{4l\beta\sqrt{\pi\hbar}}
\left\{
\left(e^{-\frac{(P-p_n)^2}{\hbar \beta^2}} 
      +e^{-\frac{(P+p_n)^2}{\hbar \beta^2}}\right) \;
\left(\Phi(\frac{z'_1}{\sqrt{\hbar}}) + 
      \Phi(\frac{z'_2}{\sqrt{\hbar}})\right)  
\right.
\nonumber \\
&+&
\left.
2\; e^{-\frac{(P^2+p_n^2)}{\hbar \beta^2}} \;
\Re e\left[ e^{-\frac{i2p_nQ}{\hbar}} 
       \left(\Phi(\frac{z'_3}{\sqrt{\hbar}}) + 
                       \Phi(\frac{z'_4}{\sqrt{\hbar}})\right) \right] 
\right\}
\; ,
\eea
where $z'_1=\beta (\frac{l}{2}+Q)$,
$z'_2=\beta (\frac{l}{2}-Q)$,
$z'_3=\beta (\frac{l}{2}+Q)-ip_n/\beta$ and 
$z'_4=\beta (\frac{l}{2}-Q)+ip_n/\beta$. 
In this case we can replace each error function by unity only in the
central region (between the vertical lines through 
$Q_{\mbox{\tiny I}}=-l/2 + p_n/\beta^2$ 
and $Q_{\mbox{\tiny II}}=l/2 - p_n/\beta^2$ in FIG.\ref{Fig2}{\bf (c)}) 
and hence recover (\ref{hurecbox}). 
Therefore, we also recover the position of the $k$'th zero along the axis,
\beq
Q_k=(2k+1)\frac{\pi\hbar}{2p_n}
\; ,
\eeq
according to (\ref{hupebox}), provided that 
$Q_{\mbox{\tiny I}} < Q_k < Q_{\mbox{\tiny II}}$.
In FIG.\ref{Fig3} we compare, on the $Q$-axis, the numerical computation
of the Husimi function (\ref{hubox}), the asymptotic approximation based 
on (\ref{hubox}) with (\ref{errofunp}) and our approximation based on the
simplified Wigner function (\ref{hupboxn}).

The form of (\ref{hurecbox}) indicates why the Husimi zeroes are
lineary distributed inside the energy curve (in this case, the phase
space box with $|p_n|=\sqrt{2mE}$).  
Indeed, the term with the hyperbolic cosine takes its lowest value 
along the $Q$-axis, which coincides with the amplitude of the 
oscillatory cosine term. 
Away from $P=0$, the hyperbolic cosine dominates the sum, descending to
a {\it valley} along this axis. 
The zeroes along the valley are determined by the minimum
value of the cosine.
The valley is very shallow because the Husimi function
decays exponentially away from the classical region, but
we can still evaluate its local minima.
The order for the spacing of zeroes is $O(\hbar)$, given by
the phase of the cosine term.

As autonomous 1-D systems always have integrable classical dynamics,
the zeroes of the Husimi functions lie over lines, as suggested
in \cite{lebvor}. 
That these lines, inside the energy curve, are {\it valleys} of 
the Husimi function is a general characteristic of these systems, 
as we will see in the following sections. 
It also seems to be a general characteristic that, when the number 
of zeroes is great, these {\it valleys} bifurcate for bounded states 
in systems where the curve of constant energy is closed. 
For fixed energy, in systems with bounded states, it is expected that
the semiclassical approximations works well for large quantum numbers.
As the number of zeroes of the Husimi function grows with the
quantum number \cite{voros1}, these bifurcations should typically
appear in the semiclassical regime.     
These bifurcations seem also to appear close to the energy curve.

For the box, the {\it valley} bifurcates close to the points 
$Q_{\mbox{\tiny I}}$ and $Q_{\mbox{\tiny II}}$ of the
$Q$-axis (FIG.\ref{Fig2}). 
Our approximation (\ref{hupboxn}) describes these bifurcating 
{\it valleys} , althought without any oscillations to indicate the
presence of zeroes (see plot ({\bf c}) of FIG.\ref{Fig2}).  
The absence of zeroes in these {\it valleys} shows that the expression
(\ref{humimic}) for the Wigner function close to the edges of the box
is not valid.
In fact, the Wigner function (\ref{wigcaja}) decreases to zero close to the 
edges of the box and is strictly zero over them. 
In constrast, the expression (\ref{humimic}) does not decrease in the 
direction of the $Q$-axis and does not vanish over the edges. 

We stress that the only approximation used to obtain (\ref{hupboxn})
is to take (\ref{humimic}) as the Wigner function. 
Moreover, by extending the limits of integration to infinity,
in the smoothing of the Wigner function, we obtain 
the expression (\ref{hurecbox}) for all points inside the box.
This approximation has no bifurcating {\it valleys} at all. 
But, as we knew, this approximation is not valid close to the edges,
because making the limits of integration go to infinity is equivalent to 
making the error functions in (\ref{hupboxn}) approach unity,   
which is not valid in the region where the bifurcations occur.


%
%

\section{Geometrical approach.}

\label{SectionGA}

In the last example we obtained an approximation to the Husimi function
by performing the Gaussian smoothing over an expression that represents 
the skeleton of the Wigner function in the semiclassical limit (\ref{humimic}). 
It was shown that this provides the general behavior inside the energy 
curve and allows us to obtain the distribution of the Husimi zeroes, 
although not close to the edges of the box. 
Here we implement a similar approach for the Husimi function of energy
eigenstates in systems where Berry's semiclassical approximations 
for the Wigner function are valid (see Section \ref{SectionRWH}).

The ideal semiclassical approximation to the Wigner function used 
in the smoothing should be Berry's {\it uniform approximation}, 
that represents the oscillations inside the energy curve $\cal E$,
and is uniformily valid along it. 
However, the integration would be very difficult to handle.
The skeleton of the {\it uniform approximation} essentially consists of an
Airy peak close to the curve $\cal E$, that in the classical limit turns 
out to be a delta function (\ref{wigclalim}) along it, and oscillations 
inside that  are well represented for the Berry's {\it simple approximation}
(\ref{berrsimple}) away from $\cal E$. 
Semiclassically, as far as integration is concerned the Airy peak 
is equivalent to the delta function.
So, for evaluation points, ${\bf X}$, close to the energy curve, 
the Husimi function is well represented by the integral,
\beq
\label{INTHUS}
I({\bf X})=
\frac{1}{2\pi^2\hbar(\omega^{-1})}\;
\int d{\bf x} \;
\delta[\mbox{H}({\bf x})-\mbox{E}] \;
\exp\left\{ -\frac{1}{\hbar}\|({\bf x}-{\bf X})\|_{\beta}^2 \right\}
\; .
\eeq
This integral is everywhere positive, smooth, and peaked along the energy 
curve $\cal E$. 
Evidently, this integral is dominated by the region where 
$\|({\bf x}-{\bf X})\|_{\beta}^2$ is a minimum: approximately
$\frac{D}{\sqrt{2\pi\hbar}}\exp\left\{ -\frac{1}{\hbar}
\|({\bf x_c}(\beta)-{\bf X})\|_{\beta}^2 \right\}$, where
${\bf x_c}(\beta)$ is the point on $\cal E$ closest to the point
${\bf X}$ in the sense of the norm $\|(\ldots)\|_{\beta}$.
Thus, close to the energy curve, this integral is essentially the
Gaussian semiclassical approximation around the torus, first encountered
by Takahashi \cite{taka} in a geometrical approach, and rederived by
Kurchan {\it et. al.} \cite{kls} in the context of the Bargmann 
representation.
These approximations have no oscillations to indicate the presence 
of zeroes. 
So, placing the evaluation point far from $\cal E$, we add to the 
integral of the delta function, a local integral over the 
{\it simple approximation}.
This is in the spirit of expression (\ref{humimic}) for the problem
of a particle in a box. 
Indeed, the only difference is that the cosine oscillations are now 
spread within the interior instead of concentrated as a delta function 
along the $q$-axis because of the particular torus geometry. 

The integration (\ref{hugausmmoth}) over the {\it simple approximation},
inside $\cal E$, can be performed analytically by making some further
approximations (for the details see Appendix \ref{OURAPP}).
The Gaussian function in (\ref{hugausmmoth}) defines an effective area 
for the integration centered on ${\bf X}$. 
This effective area can be characterized as the area of value 
$2\pi\hbar$, enclosed by the ellipse,
\beq
\label{elipse}
\frac{1}{\hbar}
\|({\bf x}-{\bf X})\|_{\beta}^2=2
\eeq
So, inside this area we can approximate the action in (\ref{berrsimple}) 
by
\beq
\label{appcenact}
S({\bf x})  \approx S({\bf X}) + {\bf \xi}({\bf X})\wedge({\bf x}-{\bf X})
\; ,
\eeq
in the semiclassical limit.
Since the denominator in (\ref{berrsimple}) does not depend on $\hbar$,
we take the simplest approximation,
\beq  
\label{appdenom} 
D({\bf x}) \approx D({\bf X})
\; .                               
\eeq
Then, the result for our approximation to the Husimi function is,
\beq
\label{SCLHUI}
H_{SCL}({\bf X}) = 
\frac{2}{\pi\sqrt{2\pi\hbar}(\omega^{-1})}
\frac{
\exp\{-\|{\bf \xi}({\bf X})\|_{\beta}^2 /4\hbar\}
      }{ 
\sqrt{D({\bf X})} 
        } 
\cos
\left\{
\frac{S({\bf X})}{\hbar} - \frac{\pi}{4} 
\right\} + 
I({\bf X})
\; .
\eeq
As we found in the last section, the Husimi zeroes inside the energy curve 
are located on a {\it valley} for 1-D systems.
The approximation (\ref{SCLHUI}) contains all the geometrical ingredients 
to understand the origin of this {\it valley}. 
This expression generally has minima rather than zeroes
in the {\it valley}.
These minima approach the Husimi zeroes when $\hbar \rightarrow 0$.
However, some of these minima could be negative in this approximation.
This problem can be fixed if we include the second order approximation,
$\frac{1}{2}\;({\bf x}-{\bf X})\,{\cal H}\,({\bf X})({\bf x}-{\bf X})^t$,
in the expansion for {\it center} action in (\ref{appcenact}),
where the Hessian matrix is
\beq
\label{hessmatrix}
{\cal H}({\bf X})=
\left[
\matrix{\partial_{qq}^2S=\partial_q(\xi_p) &
                           \partial_{qp}^2S=\partial_p(\xi_p)  \cr
        \partial_{pq}^2S=-\partial_q(\xi_q)&
                           \partial_{pp}^2S=-\partial_p(\xi_q) \cr}
\right]
\eeq
(we have applied the relations (\ref{derivs}) for the gradient of
$S({\bf x})$) and $t$ denotes the transpose.
Hence, our refined approximation 
\bea
\label{SCLHUC}
H_{SCL}({\bf X}) &=& 
\frac{2}{\pi\sqrt{2\pi\hbar}(\omega^{-1})}
\frac{
\exp\{-\Theta({\bf X}) /4\hbar\}
      }{ 
\sqrt{D({\bf X})|\det{\cal A}({\bf X})|} 
        } 
\times \nonumber \\
&\times&
\cos
\left\{
\frac{S({\bf X})}{\hbar} - \frac{\pi}{4} +
\frac{\Phi({\bf X})}{4\hbar} - 
\frac{\arg[\det{\cal A}({\bf X})]}{2} 
\right\} + I({\bf X})
\; .
\eea
Here, ${\cal A}({\bf X})$ is the complex matrix,
\beq
\label{cpmlxmatrix}
{\cal A}({\bf X})=-
\left[
\matrix{ \beta^2  &  0    \cr
          0     & 1/\beta^2 \cr}
\right] + \frac{i}{2}{\cal H}({\bf X})
\; ,
\eeq
the argument of the exponential is
\beq
\label{theta}
\Theta({\bf X})=
\frac{
\|{\bf \xi}({\bf X})\|_{\beta}^2 \;
(1-\det{\cal H}({\bf X})/4) +
\frac{1}{2} \;
{\bf \xi}({\bf X}) \,{\cal H}({\bf X})\, {\bf \xi}^t({\bf X})\;
(\partial_q(\xi_p)/2\beta^2 - \beta^2 \partial_p(\xi_q)/2)
}
{
|\det{\cal A}({\bf X})|^2
}
\; ,
\eeq
and the phase in the cosine is
\beq
\label{phi}
\Phi({\bf X})=
\frac{
-\|{\bf \xi}({\bf X})\|_{\beta}^2 \;
(\partial_q(\xi_p)/2\beta^2 - \beta^2 \partial_p(\xi_q)/2) +
\frac{1}{2}\;
{\bf \xi}({\bf X}) \,{\cal H}({\bf X})\, {\bf \xi}^t({\bf X})\;
(1-\det{\cal H}({\bf X})/4)
}
{
|\det{\cal A}({\bf X})|^2
}
\; .
\eeq
If we only use the approximation to first order for the {\it center} 
action in (\ref{appcenact}), the Hessian matrix is zero, so, 
$\det{\cal A}({\bf X})=-1$, 
$\Theta({\bf X})=\|{\bf \xi}({\bf X})\|_{\beta}^2$ 
and $\Phi({\bf X})=0$ and we recover the approximation (\ref{SCLHUI}). 
For the cases of non-normalizable states the prefactors in 
(\ref{INTHUS}), (\ref{SCLHUI}) and (\ref{SCLHUC}), change according
to the definition of the normalization of this type of states 
(see Section \ref{SectionRWH} and Section \ref{SectionPCF} for
an example). 

The expressions (\ref{SCLHUI}) and (\ref{SCLHUC}) are valid only 
inside the energy curve and depend only on the properties of the 
curve $\cal E$, like the semiclassical Wigner function.
The second order approximation to the {\it center} action,
that yields our approximation (\ref{SCLHUC}), only provides small
corrections to the argument of the exponential and specially to the
phase in the cosine. 
The corrections of the phase in the cosine improve the position of the 
minima over the {\it valley} and hence the approximation to the
zeroes.

The fact that the denominator in (\ref{SCLHUI}) vanishes on the curve 
$\cal E$ is not a major problem, because only the tips of the {\it valley} 
are close to the energy curve, where the {\it simple approximation}
plus the delta function (\ref{wigclalim}) is not a good representation of
the behavior of the Wigner function.
So, our approximations (\ref{SCLHUI}) and (\ref{SCLHUC})
do not hold close to the curve $\cal E$. 
The evaluation points ${\bf x}$, of the Wigner function, that effectivelly
contribute to the smoothing (\ref{hugausmmoth}), are enclosed by the 
ellipse (\ref{elipse}) centered on the evaluation point, ${\bf X}$, 
of the Husimi function. 
We predict that the approximation based on the mimic Wigner function
breaks down wherever the ellipse enclosing ${\bf X}$ approaches
$\cal E$.
The shape of the ellipse depends on the Husimi parameter $\beta$, so
that the region of validity of the geometrical approximation will be 
parameter dependent.

We now discuss the fact that the geometrical approximations 
(\ref{SCLHUI}) and (\ref{SCLHUC}) contain the contribution of a single
{\it chord}, even though the points within the Wigner caustic $\cal L$
are the centers of three {\it chords}.
This is simply due to the Gaussian dependence on the {\it chord} length,
defined in (\ref{norma}), which allows us to keep only the shortest 
{\it chord} in the Husimi function. 
Furthermore, we need not consider the caustic itself, because, 
as we will see, the {\it valley} of zeroes is not affected by it.
Even though the {\it simple approximation} breaks down along it, by 
predicting a spurious singularity, the correct finite Airy peak
along this line does not counterbalance the fact that the coalescing 
{\it chords} responsible for this catastrophe are longer than the normal 
third {\it chord}.
This is because the curve $\cal L$ is a locus of maximal {\it chords}. 
Therefore the Husimi function will be dominated by the single normal
{\it chord} on $\cal L$, which generates cosine oscilations well
described by the {\it simple} theory.
Hence, the sum over the differents {\it chords}, centered on points on 
the Wigner's {\it caustic} $\cal L$ and inside it, that appears in the 
{\it simple approximation} (\ref{berrsimple}), is not necessary in
(\ref{SCLHUI}) and (\ref{SCLHUC}). 
We only need to consider the {\it chord} that is continuous through
each of the two sides of $\cal L$ that are crossing the {\it valley}.
For this {\it chord} the denominator $D({\bf X})$ does not diverge
(see FIG.\ref{Fig1}).   
   
In the next sections we present two examples to show how (\ref{SCLHUC})
and (\ref{SCLHUI}) operate. The first example is an unbounded
problem whose convex energy curve is open and without a Wigner
{\it caustic}.
In this problem most of the calculations can be made analytically.
The second is a bounded problem with a closed energy curve that 
is smooth and convex. 
This example has a Wigner {\it caustic}.
In this case all the calculations were numerically.



\section{Particle subject to a constant force.}

\label{SectionPCF}

Let us apply the approach described in the last section to the problem of a
particle under the action of a constant force, $F$, that is to say with
the classical Hamiltonian $\mbox{H}({\bf x})=p^2/2m - Fq$. 
This is an unbounded problem with continuous energy spectrum where the eigenfunctions can be normalized to a delta function in $E$
({\it i.e.},
$\int \langle \psi_{E^{'}}|q \rangle\langle q|\psi_E \rangle dq=
\delta(E^{'}-E)$) \cite{baljen},  
\beq
\label{eigenctef}
\langle q |\psi_E \rangle =
\frac{1}{|F|^{1/2}}
\left[ \frac{2m|F|}{\hbar^2} \right] ^{1/3} 
\mbox{Ai}\left\{ -(q-q_r) 
\left[ \frac{2mF}{\hbar^2} \right] ^{1/3}
\right\}\; ,
\eeq
$q_r=-E/F$ is the turning point of the classical trajectory for 
an energy $E$. 
The Wigner function is in this case \cite{balaz,baljen}
\beq
\label{wigfuerza}
W({\bf x})=
\left[\frac{8m}{\hbar^2 F^2}\right]^{1/3}
\mbox{Ai}\left\{
\left[\frac{8m}{\hbar^2 F^2}\right]^{1/3}
(\mbox{H}({\bf x})-E)
\right\}.
\eeq
It is easy to see that this expression coincides with (\ref{berrtran})
(the prefactor equal to unity in this unbounded problem, according to the
normalization choosen above).
So, the {\it Transitional approximation} to the Wigner function is exact
in this case.

The Husimi function for this problem can be caculated analytically 
(Appendix \ref{HFC}), the result is,
\bea
\label{husfuerza}
H({\bf X})&=&|B|^2 
\exp
\left\{
-\frac{1}{\hbar}
\left(
\frac{P^2}{\beta^2} + \frac{2mF}{\beta^2}\;Q
\right)
\right\} \times \nonumber \\
\label{husimif}
&\times&
\left|
\mbox{Ai}
\left\{
-\left(
Q - Q_* - i P/\beta^2
\right)
\left[\frac{2mF}{\hbar^2}\right]^{1/3}
\right\}
\right|^2
\; ,
\eea
where $Q_*=q_r + mF/2\beta^4$ and $|B|^2$ is the normalization constant. 
The distribution of zeroes, in the concave side
of the curve $\cal E$, of constant energy, is shown in FIG.\ref{Fig4}.
Since the zeroes are those of the Airy function in (\ref{husfuerza}), which
only occur  for a negative real argument, their distribution 
is along the $Q$-axis for any $\hbar$ value. 
For an energy $E$ the classical turning point is fixed and so is 
$Q_*$. 
Thus, we only have to change the scale for the ${\bf X}$ coordinates 
to make the Airy function in (\ref{husfuerza}) invariant with $\hbar$. 
Due to this scaling property, we can analyze semiclassically the 
distribution of zeroes by fixing $\hbar$ and by looking at the 
behavior of the Airy function for values of $Q$ far away from $Q_*$ 
on the concave side of $\cal E$.
In this region we can replace the Airy function in (\ref{husfuerza}) 
by its asymptotic form (\ref{airyapp}), where now the argument is 
complex.
The result is the same if we construct the Husimi function 
in (\ref{hubarg}), with the approximation to the Bargmann function
(\ref{bargfunwkb}), obtained in Appendix \ref{CWKBM} by the complex 
WKB method.
Therefore, the direct semiclassical analysis of expression 
(\ref{husfuerza}), or the application of the complex WKB method, bring 
about the same distribution of the zeroes in the concave side of 
$\cal E$. This distribution is given by the zeroes of the cosine in
(\ref{bargfunwkb}) over the real axis, 
\beq
\label{zeroeswkb}
Q_k=\frac{[3\pi\hbar(2k+3/2)]^{2/3}}{2^{5/3}(mF)^{1/3}}
+ Q_*
\hspace{2cm} 
k=0,1,2,\ldots
\;\;\; ,
\eeq
where now $F>0$. 
The order for the spacing between the $k$'th and the $(k+1)$'th zero 
is $O(\hbar^{2/3})$.
However, these zeroes accumulate on $Q_{*}$ as $\hbar \rightarrow 0$,
so it is more relevant to derive the asymptotic spacing near a fixed
position $Q$. 
Approximating the Airy function in (\ref{husfuerza}) by its large 
argument from (\ref{airyapp}), we then obtain the spacing of minima
as $O(\hbar)$ in agreement with Leboeuf and Voros \cite{lebvor}.

To compare these results with the geometrical approximation, we note that 
for the particle subject to a constant force, the {\it centre} action is
\beq
S({\bf X})=\frac{2}{3}\left(\frac{8m}{F^2}\right)^{1/2}
\;[-(H({\bf X})-E)]^{3/2}
\eeq
and the denominator in (\ref{berrsimple}) is
\beq
D({\bf X})=\frac{F}{m}\;\;\xi_p=
\left(
\frac{8F^2}{m}
\right)^{1/2}
[-(H({\bf X})-E)]^{1/2}
\; .
\eeq
Hence, in our approximations (\ref{SCLHUI}) and (\ref{SCLHUC})
the first term can be calculated analytically, where now the prefactor is
$(8/\pi\hbar)^{1/2}$.
The classical limit (\ref{wigclalim}) can be obtained by applying formula 
(\ref{clalim}) to the Wigner function (\ref{wigfuerza}). 
The integral (\ref{INTHUS}) over $\cal E$ becomes,
\beq
\label{intfc}
I({\bf X})=\frac{1}{\pi\hbar |F|}
\int dp 
\exp\left\{
-\frac{1}{\hbar}[\beta^2(q_{\tiny E}(p)-Q)^2+(p-P)/\beta^2]
\right\}
\eeq
where $q_{\tiny E}(p)=p^2/2mF + q_r$.
This is a non-oscillatory smooth function, peaked on $\cal E$, that
decreases monotonically away from the energy curve.

The geometrical origin of the {\it valley} of zeroes along the axis $P=0$, 
in the concave side of $\cal E$, can now be understood with the help of our
approximation (\ref{SCLHUI}).  
In fact, from the argument of the exponential, we see 
that for values of $\beta$ allowed by our approximations 
(see Section \ref{SectionGA}), $\|{\bf \xi}({\bf X})\|_{\beta}^2$ 
is essentially equal to the square of the 
{\it chord}'s length. 
Along the $Q$-axis, FIG.\ref{Fig5} shows that the prefactor of 
the cosine in (\ref{SCLHUI}) has almost the same value as the 
integral (\ref{intfc}). 
Away from this axis, the length of the {\it chord} grows, making 
the oscillatory term so small that the second term dominates the sum, 
creating in this way a {\it valley}. 
Along this {\it valley} the oscillations of the cosine generate a 
sequence of local minima. 
However, the position of these local minima are shifted, relative to the
position of the Husimi zeroes, approximately by the distance $(Q_{*}-q_r)$
(FIG.\ref{Fig6}). 
Moreover, for points $Q$ far away from $Q_{*}$, these local minima 
become negative because the prefactor of the cosine becomes greater 
than the integral (see FIG.\ref{Fig5}). 

The corrections given by our second order approximation (\ref{SCLHUC}) 
fix these problems. 
In fact, the corrections to the argument of the exponential, given by
(\ref{theta}), ensure that the local minima are positive on the axis
and the corrections to the phase in the cosine, given by (\ref{phi}),
improve the position of the minima relative to the Husimi zeroes 
(see FIG.\ref{Fig5} and FIG.\ref{Fig6}).
We also compare, in FIG.\ref{Fig6}, the general behavior  
of (\ref{SCLHUC}) and the Husimi function calculated numerically
on $P=0$.
One observes a general agreement that improves as $Q$ recedes
from $Q_{*}$ (equivalent to the limit $\hbar \rightarrow 0$, in this
problem). 
In this semiclassical limit, the same figure shows that the minima 
become zeroes. 
FIG.\ref{Fig7} shows the relative error between 
the position of the Husimi zeroes (calculated numerically),
the position being given by the minima of (\ref{SCLHUC}) and 
(\ref{SCLHUI}) (shifted by the distance $(Q_{*}-q_r)$) and the zeroes
(\ref{zeroeswkb}). As expected, our approximation (\ref{SCLHUC})
does not work well close to the energy curve where the mimic of the
Wigner function used for the smoothing is not a good approximation.
   


\section{Generic case.} 

\label{SectionGCQO}

In this section we apply our geometrical approach to the problem of
a particle subject to an asymmetric anharmonic potential.
The classical Hamiltonian is
\beq
\label{hamilanhar}
\mbox{H}({\bf x})=\frac{p^2}{2m} + 
\frac{m\omega_0^2}{2}\;(q-q_0)^2 +
\frac{\lambda}{2}\;q^4
\; .
\eeq
This system is an example of a general system with a convex, closed
energy curve ${\cal E}$ and a Wigner caustic ${\cal L}$. 
FIG.\ref{Fig1} shows typical curves, ${\cal E}$ and ${\cal L}$, 
in this system.

In this bounded problem, we fixed the classical energy curve at 
$E\approx 30.8175$ and we calculated numerically the distribution 
of the zeroes of the Husimi functions inside it for two energy 
eigenstates.
The latter correspond to two values of quantized $\hbar$:
the eigenstate $n=30$ for a value of $\hbar\approx 0.508236$, and 
the second is the eigenstate $n=45$ for a value of 
$\hbar\approx 0.340691$.
In order to simplify the calculations, all the others parameters of the
problem (including $\beta$) were set to unity, except $q_0=4.0$ and 
$\lambda=0.1$.
 
FIG.\ref{Fig8} shows the distribution of the Husimi zeroes for these 
states.
The zeroes are distributed along lines, as expected for a system with 
integrable classical dynamics \cite{lebvor}.
These lines are very shallow {\it valleys}  of the Husimi function.
Since the energy curve is symmetric with respect to the $Q$-axis, the
distribution of zeroes also mantains this characteristic.   
In Section \ref{SectionPB} we anticipated, as a general charcteristic,
the bifurcation of the {\it valleys} in the semiclassical regime for
bounded states in systems with a closed energy curve.
Here, we have a generic example where a principal {\it valley} bifurcates 
in each half plane (FIG.\ref{Fig8}).
The asymmetry in the lengths of these bifurcating {\it valleys} 
reflects the asymmetry of the curve $\cal E$ with respect to the
$P$-axis. 
For each quantum number $n$, the majority of the Husimi zeroes 
belong to the principal {\it valley}.
For the parameter $\beta$ chosen, the principal {\it valley} runs 
parallel to the vertical side of the caustic.
At the middle it passes very close on the outside of $\cal L$ and
them crosses the cusps of the caustic (see FIG.\ref{Fig8}). 
This shows that the distribution of the zeroes
is not affected by the presence of the Wigner caustic.

Our approximation (\ref{SCLHUI}) to the Husimi function inside $\cal E$,
supplies the geometrical insight for the origin of the principal 
{\it valley}.   
As we saw in the example of the last section, this {\it valley}
is located where both terms of the approximation are of the same order.
The amplitude of the oscillatory term is dominated by the exponential,
whose argument is essentially equal to the square of the {\it chord}'s
length. 
If we follow the {\it chord}'s length along the level curves of the 
{\it center} action, that is, essentially the phase curves of the cosine 
in the oscillatory term, we observe local minima of this length restricted 
to these curves.
Away from the {\it valley} the length of the {\it chords} grows, 
making the oscillatory term so small that the smooth second term 
dominates the sum. 
This is also what happend in the problem of the constant force
(Section \ref{SectionPCF}) where the $Q$-axis is the locus of minima of 
the {\it chord}'s length restricted to the level curves of the {\it center} 
action which cross the axis orthogonally.
Along the {\it valley}, the oscillations of the cosine produce a serie
of local minima of (\ref{SCLHUI}) that indicate, in a first
approximation, the position of the Husimi zeroes.
The position of these minima are very close to the points where the cosine
takes its minimum value, only slightly modified when we consider
the sum of the two terms.
In FIG.\ref{Fig9} {\bf (a)} and FIG.\ref{Fig10} {\bf (a)}
we illustrate, for each quantum number, the geometrical method to locate
the {\it valley} and the Husimi minima. 
We display the level curves of the {\it center} action for which
the cosine takes its lowest value, and the level curves of the 
{\it chord}'s length. 
The tangency of the two sets of curves determine the restricted minima
of the {\it chord}'s length along the chosen level curves of the 
{\it center} action.
These points are candidates to be, approximately, the local minima of
(\ref{SCLHUI}) after performing the sum of the two terms.
The {\it valley} of minima passes through all these points in this
approximation.
The comparison in FIG.\ref{Fig8} between the {\it valley} of minima of 
(\ref{SCLHUI}) and the zeroes of the Husimi function shows that our approximation to the principal {\it valley} of zeroes holds farely well
until the bifurcation. 
Although this approximation to the {\it valley} continues
after the bifurcation, it does not take into account the bifurcation itself.
Even this continuation, is no longer such a good approximation of the longer
{\it valley} beyond the bifurcation.
 
The local minima of (\ref{SCLHUI}) along the principal 
{\it valley} have almost the same spacing as the Husimi zeroes.
However, the absolute position of the predicted zeroes is not so precise
(see FIG.\ref{Fig9} {\bf (c)} and FIG.\ref{Fig10} {\bf (c)}).
Furthermore, some of these minima become negative, because the prefactor in 
the cosine becomes greater than the integral in the second term of the
approximation. 
We found the same situation when we applied (\ref{SCLHUI}) in the last 
section, so we again make use of the refined approximation (\ref{SCLHUC}).
This supplies corrections to the {\it chord}'s length in the argument of
the exponential and to the {\it center} action in the phase of the cosine. 
Therefore, we can use the same geometrical method to find the approximate
position of the local minima of (\ref{SCLHUC}) of the Husimi zeroes in 
the semiclassical regime. 
Hence, FIG.\ref{Fig9} {\bf (b)} and FIG.\ref{Fig10} {\bf (b)} display
the phase curves of the cosine for minimum values, and the level curves
of the argument of the exponential. 
The tangencies of these two sets of curves determine the restricted minima
of the argument of the exponential in (\ref{SCLHUC}), along the level curves
of the phase of the cosine.
Clearly, these points belong to a {\it valley} because, away from the line 
that passes through all the restricted minima, the smooth second term 
in (\ref{SCLHUC}) dominates the sum exponentially.
Moreover, since the cosine takes its lowest value at these points, they
are close to the local minima of (\ref{SCLHUC}).
The comparison of these points with the Husimi zeroes in FIG.\ref{Fig9}
{\bf (d)} and FIG.\ref{Fig10} {\bf (d)}, shows that they are a good
approximation to the zeroes along the principal {\it valley}
until the bifurcation.

Our approximation (\ref{SCLHUC}) also fails to take account of the 
bifurcation.
The {\it valley} of local minima is again a very good approximation to 
the principal {\it valley} of the Husimi function, but now, it   
also represents accurately its continuation along the longest bifurcating
{\it valley} (see FIG.\ref{Fig8}).
However, along this bifurcating {\it valley} there are no
local minima of (\ref{SCLHUC}) to indicate the presence of zeroes, 
because there the prefactor of the oscillatory term is much smaller 
than the integral of the second term.
This can be observed in FIG.\ref{Fig11} and FIG.\ref{Fig12} where
we display the logarithm of our approximation and the Husimi function
along the {\it valley} of local minima of (\ref{SCLHUC}), for each
quantum number.

We saw that the {\it valley} of zeroes of the Husimi function is not
affected by the Wigner caustic. 
The principal {\it valley}, that runs parallel to the vertical side of
the caustic on the outside of $\cal L$ , crosses the cusps of the caustic
at the tips of this side. 
This is reflected in our approximations (\ref{SCLHUI}) and (\ref{SCLHUC})
since only the contribution of a single {\it chord} inside the Wigner caustic
is needed.
Because of the dependence on the {\it chord}'s length in our approximations,
we showed that we only need to consider the shortest {\it chord} that is
continuous through $\cal L$, irrespective of the two {\it chords} that 
coalesce along this caustic.
This shortest {\it chord} generates the {\it valley} that runs parallel
and very close to the principal {\it valley} of the Husimi
function unaffected by the close lying Wigner caustic (see FIG.\ref{Fig8}).

To end this section, we note that the consideration of states corresponding
to a fixed energy, that are quantized by varying $\hbar$, leads to a 
spacing of the Husimi zeroes of $O(\hbar)$.
This is easily seen in our simple approximation (\ref{SCLHUI}), for a 
fixed location along the {\it valley} that is classically determined.
The wavelength of the oscillations that determine the minima
is proportional to $\hbar$, whereas $S({\bf x})$ is a classical action.

%
%
%

\section{Conclusions.}

\label{Conclusions}

The semiclassical approximation of the Husimi function has been derived
by integrating the semiclassical Wigner function with a Gaussian window.
This is not fundamentally different from the calculation of probability
densities of position or momenta as projections of the Wigner function,
except that now we project onto coherent states.
In each projection, we obtain a classical approximation by substituting
the Wigner function by a delta-function along the classically allowed
region.
Here, this leads to a narrow ridge along the classical region, which
is supplemented by an oscillatory term derived from the {\it centre} 
and {\it chord} structure within the energy curve.
The oscillations of the latter along the classical shallow {\it valleys}
combine to form local minima, that indicate the positions of the Husimi
zeroes inside the energy curve.
This geometrical explanation of the linear distribution of minima
cannot be pushed to the point of predicting absolute zeroes, but it
may be nonetheless surprising that their positions are asymptotically
accurate, though obtained by subtracting two exponentially small
terms.

The advantage of deriving the intermediate approximation (\ref{SCLHUI})
is that the location of the {\it valley} has a simple dependence on the
minimal {\it chord} along curves of constant {\it centre} action 
$S({\bf x})$. This curve is purely classical, once the excentricity 
$\beta$ of the coherent states defines the phase space metric. 
The corrections added to our complete formula (\ref{SCLHUC}) are also 
classical.
They mostly alter the distribution of zeroes along the {\it valley}, 
so we find that the {\it valleys} are basically determined by the 
classical structure, in agreement with Leboeuf and Voros \cite{lebvor}.

It is perhaps surprising that the Wigner caustic $\cal L$ does not affect 
the position of the Husimi zeroes. 
However, this fact is in agreement with previous calculations for the 
projections of the Wigner function \cite{ozo82}.
In each case the integration singles out a contributing {\it chord} 
from the semiclassical Wigner function, while ignoring the other 
possibly singular {\it chords}. 
Thus we can understand the complexity of the Wigner function as arising
from the necessity to account for diverse square integrable projections.

We have limitted our considerations to autonomous Hamiltonian systems
with one degree of freedom, which are necessarily integrable.
Our approximations are equally valid for integrable classical maps
on the plane onto itself: the zeroes are always predicted to lie
along the locus of minimal {\it chords}, the minimum being evaluated
along lines of constant phase for the Wigner function. 

The {\it chord} structure generalizes to tori of higher dimension
\cite{ozha}, so our methods will also be extendable to the study 
of Husimi functions of integrable systems with more than a single
degree of freedom.
In particular they may help to define the zero-manifolds in this case.
For chaotic systems, we know that the {\it chord} structure is
also present, though it involves individual orbits \cite{alfredo}.
The challenge lies open, to explore the relation between the 
structure of the Husimi and the Wigner functions for nonintegrable
systems.

%
%

\acknowledgements
Discussions with R.O. Vallejos, M. Saraceno and A. Voros are gratefully acknowledged.
FT thanks CLAF-CNPq for financial support, as well as the overall support
of Pronex-MCT.

\appendix

%
%

\section{}

\label{HFPB}

{\it The Husimi function for the problem of a particle in a box 
with hard walls.}

\vspace{1pc}
Here we show the fundamental steps in the derivation of (\ref{hubox}). 
We start with the expression of the Husimi function given by the formula
(\ref{husdef1}),
\beq
\label{husappa}
H({\bf X})=\frac{1}{2\pi\hbar}
\left|
\int_{-\infty}^{+\infty}
 \langle \Omega_{\bf X}|   q  \rangle
 \langle        q      | \psi_n \rangle
dq
\right|^2 
=
\frac{\beta}{l(\pi\hbar)^{3/2}}
\left|
\int_{-l/2}^{+l/2}
e^{-\frac{\beta^2}{2\hbar}(q-Q)^2-i\frac{Pq}{\hbar}}
\cos(p_nq/\hbar)
\right|^2
\;,
\eeq
where $\langle q | \Omega_{\bf X} \rangle$ is the normalized
coherent states in the position representation (see for example 
\cite{cohen}) and $\langle q | \psi_n \rangle$ the even eigenfunction
(\ref{boxeigen}). 
If we express the cosine in the last integral as
$(1/2)(e^{ip_nq/\hbar}+e^{-ip_nq/\hbar})$ we have
\bea
H({\bf X})&=&
\frac{\beta}{4l(\pi\hbar)^{3/2}}
\left| 
e^{-\frac{(P-p_n)^2}{2\hbar\beta^2}}
\,
e^{-i\frac{(P-p_n)Q}{\hbar}} 
\int_{-l/2}^{+l/2}
e^{
-\left\{
\frac{1}{\sqrt{2\hbar}}
[\beta(q-Q)+i(P-p_n)/\beta]
\right\}^2
}
dq
\;
+ 
\right.
\nonumber
\\
&&
+
\left.
e^{-\frac{(P+p_n)^2}{2\hbar\beta^2}}
\, 
e^{-i\frac{(P+p_n)Q}{\hbar}} 
\int_{-l/2}^{+l/2}
e^{
-\left\{
\frac{1}{\sqrt{2\hbar}}
[\beta(q-Q)+i(P+p_n)/\beta]
\right\}^2
}
dq
\right|^2
\; .
\eea
Making a change of variables in the expression between 
braces in each integral and using the definition of the error function
\beq
\Phi(w)=\frac{2}{\sqrt{\pi}}
\int_{0}^{w}
e^{-y^2}
dy
\; ,
\eeq
the Husimi function becomes
\bea
H({\bf X})&=&
\frac{1}{8l\beta\sqrt{\pi\hbar}}
\left| 
e^{-\frac{(P-p_n)^2}{2\hbar\beta^2}}
\,
e^{-i\frac{(P-p_n)Q}{\hbar}} 
\Phi(\frac{z_1}{\sqrt{2\hbar}}) + \Phi(\frac{z_2}{\sqrt{2\hbar}})
\right.
\nonumber
\\
&&
+
\left.
e^{-\frac{(P+p_n)^2}{2\hbar\beta^2}}
\, 
e^{-i\frac{(P+p_n)Q}{\hbar}} 
\Phi(\frac{z_3}{\sqrt{2\hbar}}) + \Phi(\frac{z_4}{\sqrt{2\hbar}})
\right|^2
\; .
\eea
With the help of the identity 
$|w_1+w_2|^2=|w_1|^2+|w_2|^2+\,2\,\Re e(w_1w_2)$ of the complex numbers 
we arrive to the expression (\ref{hubox}) for the Husimi function in this
problem.

Finally, we give the asymptotic expansion of the error function
$\Phi(w)$, used in Section \ref{SectionPB}, in view of the confused 
and even incomplete form that it appears in the usual references 
\cite{abramo,gradry},
\beq
\Phi(w) \approx
\left\{
\matrix{ 1-\frac{e^{-w^2}}{\sqrt{\pi}w}[F_n(w)+O(|w|^{-2(n+1)})]&
                                                             \Re e(w) > 0 \cr 
        -1-\frac{e^{-w^2}}{\sqrt{\pi}w}[F_n(w)+O(|w|^{-2(n+1)})]& 
                                                             \Re e(w) < 0 \cr}
\right.
\;\;\;\;|w|\;\; \mbox{large}
\eeq
where $F_n(w)= \sum_{k=0}^{n}\frac{(-1)^k(2k-1)!!}{(2z^2)^k}$.

%

\section{}

\label{OURAPP}

{\it Details of the geometrical approximation to the Husimi function
within the energy curve.}

Here we derive the oscillatory term of our expression (\ref{SCLHUC}).
We start with the Gaussian smoothing (\ref{hugausmmoth}) over the
{\it simple approximation} (\ref{berrsimple}) within the energy curve.
If we apply the approximation (\ref{appcenact}) to the {\it center} action, 
incluing the second order term 
$\frac{1}{2}\;({\bf x}-{\bf X})\,{\cal H}\,({\bf X})({\bf x}-{\bf X})^t$ 
(with ${\cal H\,({\bf X})}$ the Hessian matrix (\ref{hessmatrix})),
and the approximation (\ref{appdenom}) for the denominator, we have
\beq
\label{oscterm}
\frac{2}{\pi\sqrt{2\pi\hbar}(\omega^{-1})\pi\hbar\sqrt{D({\bf X})}} \;
\Re e 
\left[
\exp 
\left\{ i
\left(
\frac{S({\bf X})}{\hbar} - \frac{\pi}{4}
\right)
\right\}
{\cal I}  
({\bf X})
\right]
\; ,
\eeq 
where ${\cal I} ({\bf X})$ is the integral,
\beq
{\cal I} ({\bf X})=
\int_{-\infty}^{+\infty} dq \int_{-\infty}^{+\infty} dp \;\;
e^{
\frac{1}{\hbar}
\left[
-a_1(q-Q)^2 - a_2(p-P)^2 + a_3(q-Q)(p-P) + a_4(q-Q) + a_5(p-P) 
\right]
}
\;\; ,
\eeq
with the complex coefficients: 
$a_1({\bf X})=\beta^2-i\,\partial_q(\xi_p)/2$, 
$a_2({\bf X})=1/\beta^2 + \,i\, \partial_p(\xi_q)/2$,
$a_3({\bf X})=i\;\partial_p(\xi_p)$, $a_4({\bf X})=i\;\xi_P$ and 
$a_5({\bf X})=-i\;\xi_Q$.
This double Gaussian integral can be performed: 
\beq
{\cal I} ({\bf X})=
\frac{\pi\hbar}
{\sqrt{|a_1( a_2 - a_3^2/4a_1)|}}
\;
\exp
\left\{
\frac{1}{4\hbar}
\left[
\frac{a_4^2}{a_1} + \frac{(a_5+a_3a_4/2a_1)^2}{(a_2 - a_3^2/4a_1)} 
\right]
-\;i\;
\frac{(\theta_1+\theta_2)}{2}
\right\}
\; ,
\eeq
where $\theta_1=\arg(a_1)$  
and $\theta_2=\arg(a_2 - a_3^2/4a_1)$ with
$\pi/2 <\theta_1,\theta_2 < \pi/2$.
This result can be written in a more elegant way with the help of the
complex matrix, ${\cal A} ({\bf X})$, defined  in (\ref{cpmlxmatrix}), 
\beq
\label{inteaux}
{\cal I} ({\bf X})=
\frac{\pi\hbar}
{\sqrt{|\det{\cal A}({\bf X})|}}
\;
\exp
\left\{
\frac{1}{4\hbar}
\frac{{\bf \xi}({\bf X})\;{\cal A}({\bf X})\;{\bf \xi}^t({\bf X}) }
{\det{\cal A}({\bf X})} 
-i\;\frac{\arg(\det{\cal A}({\bf X}))}{2} 
\right\}
\; .
\eeq
Replacing (\ref{inteaux}) in (\ref{oscterm}) and solving the real
part we find the oscillatoty term in (\ref{SCLHUC}).

%
%

\section{}

\label{HFC}

{\it The Husimi function for the problem of a particle subject to a 
constant force.}

\vspace{1pc}
The idea is to solve the Schr\"odinger equation for the eigenstate
$|\psi_E \rangle$, written in the Bargmann 
representation, to find the Bargmann function in (\ref{hubarg}) for this
problem. To write the equation in the Bargmann representation, we proceed 
in the standard form: we write the quantum Hamiltonian as a function of
the {\it creation} and {\it destruction} operators
(\ref{desoperator}) in normal order 
({\it i.e.} the $\hat{a}^{\dagger}$ operators to the right of the
$\hat{a}$ operators) and then using,      
\beq
\langle z|\hat{a}|\psi \rangle =\hbar\;\partial_z \langle z|\psi \rangle
\hspace{2cm} 
\langle z|\hat{a}^{\dagger}|\psi \rangle = z \langle z|\psi \rangle \;\;,
\eeq 
(where the $|z\rangle$ are the unnormalized coherent states
defined in Section \ref{SectionRWH}) we get 
\beq
\left\{
-\frac{\beta^2}{4m}
[\hbar^2\partial_z^2 - 2\hbar z\partial_z + z^2 - \hbar]
-\frac{F}{\sqrt{2}\beta}z-\frac{F\hbar}{\sqrt{2}\beta}\;\partial_z - E 
\right\}
\langle z|\psi_E \rangle=0 \;\;.
\eeq
This equation can be put in the form
\beq
\label{difeq}
\{[\partial_z-f(z)]^2 + c(z-z_*)\}
\langle z|\psi_E \rangle=0
\; ,
\eeq
where $c=(1/\hbar^2)8mF/\beta^3\sqrt{2}$, 
$z_*=(\beta/\sqrt{2})(q_r + mF/2\beta^4)$ 
(with $q_r=-E/F$ the turning point of the classical trajectory
of energy $E$), 
and the function $f(z)=(1/\hbar)(z-(2mF/\beta^3\sqrt{2}))$. 
The general solution of (\ref{difeq}) is
\beq
\langle z|\psi_E \rangle=
\exp\{g(z)\}\;
\mbox{Ai}\{ -(z-z_*)[c]^{1/3} \}
\hspace{0.5cm}
\mbox{with}
\hspace{0.5cm}
\partial_z g(z)=f(z)\;\;.
\eeq
Therefore, the Bargmann function for an eigenstate of the problem of 
a particle subject to a constant force is
%
\beq
\label{bargf}
\langle z|\psi_E \rangle= B 
\exp
\left\{
\frac{1}{\hbar}
\left(
\frac{z^2}{2} - \frac{2mF}{\sqrt{2}\beta^3}z
\right)
\right\}
\mbox{Ai}
\left\{
-\left(
z-z_*
\right)
\left[\frac{8mF}{\hbar^2\sqrt{2}\beta^3}\right]^{1/3}
\right\}
\; ,
\eeq
where $B$ is a complex constant.
Replacing (\ref{bargf}) in (\ref{hubarg}) we obtain the Husimi function
(\ref{husimif}).

%
%

\section{}

\label{CWKBM}

{\it A semiclassical approximation to the Husimi function for the problem
of a particle subject to a constant force through the WKB method 
in the Bargmann representation.}

\vspace{1pc}
Here we follow the WKB construction in the Bargmann representation given
by Voros \cite{voros1} to obtain a semiclassical approximation to
the Bargmann function $\langle z|\psi_E \rangle$ and then, through 
(\ref{hubarg}), a semiclassical approximation to the Husimi function.

We use the WKB construction based on the Weyl symbol $H_W({\bf x})$, of 
the quantum Hamiltonian $\hat{\mbox{H}}$, for the cases where it does
not depend on $\hbar$ ({\it i.e} that it coincides with the classical 
Hamiltonian, $\mbox{H}$).
It is easy to verify, through (\ref{asymbol}), that  
$H_W\equiv\mbox{H}$ when the classical Hamiltonian is of 
the form (\ref{hamil}). As this is the case for this problem,
we write $\mbox{H}$ instead of $H_W$. 

Voros \cite{voros1} argues that to build a WKB solution of 
the equation $\hat{\mbox{H}}|\psi_E \rangle = E |\psi_E \rangle$ in 
the Bargmann representation, we should just apply the same formulae
as for the Schr\"odinger representation ({\it i.e.} the 
common position representation), while replacing 
${\bf x}=(q,p) \rightarrow (z,\bar{z})$ and 
$\hbar \rightarrow i\hbar$. 
Therefore, the eigenvalue equation admits local asymptotic solutions 
to leading order in $\hbar$,
\beq
\label{wkbb1}
\langle z|\psi_E \rangle \approx
\left[
\frac{\partial \mbox{H}}{\partial \bar{z}}
\right]_{\bar{z}_E(z)}^{-1/2} 
\exp\{S(z)/\hbar\}
\; ,
\eeq
where $\bar{z}_E(z)$ is the function defined implicitily by the 
relation that defines the classical energy curve in the $(z,\bar{z})$
coordinates,
\beq
\label{energycurve}
\mbox{H}(z,\bar{z})=E
\; ,
\eeq
and $S(z)$ is the classical action in the complex coordinates,
\beq
S(z)=\int^z \bar{z}_E(z^{'}) dz^{'}\;.
\eeq
The fact that $\bar{z}$ and its complex conjugate $z$ are related 
themself by $\mbox{H}(z,\bar{z})=E$ restricts $z$ to the real energy 
curve. 
This means that the function $\bar{z}_E(z)$ is a branch of 
(\ref{energycurve}),  defined in a sheet to which the
energy curve belongs, and is single valued over it.
For nonanalytic $\mbox{H}$, there is no guarantee to have an analytical
continuation of $\bar{z}_E(z)$ anywhere outside. Thus (\ref{wkbb1}) is
well defined only for $z$ on the real energy curve and it is globally
regular, since there is no turning point anywhere. 

In order to obtain a holomorphic approximation far away from the energy curve
we use the fact that, in this problem, the classical Hamiltonian is analytic 
in both variables $z$ and $\bar{z}$, so the energy relation 
(\ref{energycurve}) defines implicitily $\bar{z}_E$ as a multiply 
valued function of $z$ ({\it i.e} its domain are the sheets of some 
Riemannian surface). 
However, outside the real energy curve $\bar{z}_E(z)$ is no longer the 
complex conjugate of $z$, so we use a less confusing notation denoting 
as $y$ the independent complex variable canonically conjugate to $z$,   
\beq
z=\frac{1}{\sqrt{2}}
\left(\beta q-i\frac{p}{\beta}\right)
\hspace{2cm}
y=\frac{1}{\sqrt{2}}
\left(\beta q+i\frac{p}{\beta}\right)
\; ,
\eeq
where $q$ and $p$ are now complex. Then, the complex energy curve in this 
problem is  
\beq
\label{eq2degre}
\mbox{H}(z,y)=a_0y^2+a_1(z)y+a_2(z)=0
\; ,
\eeq
with the coefficients:
$a_0=-\beta^2/4m$,
$a_1(z)=\beta^2z/2m - F/\sqrt{2}\beta$ and
$a_2(z)=-\beta^2 z^2/4m - Fz/\sqrt{2}\beta - E$. This is an equation of
degree two, so the explicit branches $y=y_E(z)$ are defined over a 
two-sheet Riemannian surface.
If we make the simple change of variable $w=2a_0y+a_1$, we obtain
the equivalent equation
\beq
w^2-u(z)=0
\; ,
\eeq
where $u(z)=a_1^2 - 4a_0a_2=-[\sqrt{2}\beta F/m](z-z_*)$ 
($z_*=[\beta/\sqrt{2}](mF/2\beta^4 +q_r)$ and $q_r=-E/F$ is the turning 
point of the classical trajectory of energy $E$). The function $w(z)$ 
is defined over a Riemann surface with branch points $z=z_*$ and 
$z=+\infty$.
The branches are 
$w_{\mbox{\tiny I}}(z)=i[\sqrt{2}\beta F/m]^{1/2}\sqrt{z-z_*}$ and
$w_{\mbox{\tiny II}}(z)=-i[\sqrt{2}\beta F/m]^{1/2}\sqrt{z-z_*}$, ($F>0$).
Here we use the notation: $\sqrt{z-z_*}=\sqrt{r}e^{i\theta/2}$, 
$0<\theta<2\pi$; so we can also consider $z$ in the ordinary complex
plane and $w_{\mbox{\tiny I}}(z)$ and $w_{\mbox{\tiny II}}(z)$ as
two different functions.
Since $y$ is a single valued function of $w$, the branches $y_E(z)$,
or equivalently the solutions of (\ref{eq2degre}) are, 
\beq
\matrix{ y_{\mbox{\tiny I}}(z)=
-i\left[ \frac{8mF}{\sqrt{2}\beta^3} \right]^{1/2} 
\sqrt{z-z_*} +z - \frac{2mF}{\sqrt{2}\beta^3} \cr
         y_{\mbox{\tiny II}}(z)=
\;\;\;i\left[ \frac{8mF}{\sqrt{2}\beta^3} \right]^{1/2}
\sqrt{z-z_*} +z - \frac{2mF}{\sqrt{2}\beta^3} \cr }
\hspace{2cm}
(F>0)\;.
\eeq

The WKB approximation is then a linear combination of solutions of the type
(\ref{wkbb1}) for each branch, valid away from the energy curve: 
\beq
\langle z|\psi_E \rangle \approx
\left[
\frac{\partial \mbox{H}}{\partial y}
\right]_{y_{\mbox{\tiny I}}(z)}^{-1/2} 
\exp\{S_{\mbox{\tiny I}}(z)/\hbar\}
+
\left[
\frac{\partial \mbox{H}}{\partial y}
\right]_{y_{\mbox{\tiny II}}(z)}^{-1/2}
\exp\{S_{\mbox{\tiny II}}(z)/\hbar\}
\;,
\eeq
with the complex actions for each branch
\beq
S_{\mbox{\tiny I}}(z)=
\int^z y_{\mbox{\tiny I}}(z^{'}) dz^{'}
\hspace{2cm}
S_{\mbox{\tiny II}}(z)=
\int^z y_{\mbox{\tiny II}}(z^{'}) dz^{'}
\;.
\eeq
%
Hence, the semiclassical approximation to the Bargmann function in
this problem is
\bea
\label{bargfunwkb}
\langle z|\psi_E \rangle 
&\approx&
B^{'}
\exp
\left\{
\frac{1}{\hbar}
\left(
\frac{z^2}{2} - \frac{2mF}{\sqrt{2}\beta^3}z
\right)
\right\} 
(z-z_*)^{-1/4}
\times \nonumber \\
&\times&
\cos
\left\{
\frac{2}{3\hbar}
\left[
\frac{8mF}{\sqrt{2}\beta^3}
\right]^{1/2}
(z-z_*)^{3/2} - \frac{\pi}{4}
\right\}
\; .
\eea
This expression can also be obtained applying the asymptotic form 
(\ref{airyapp}) to the Airy function in the Bargmann function 
(\ref{bargf}) of Appendix \ref{HFC}. 
Replacing (\ref{bargfunwkb}) in (\ref{hubarg}) leads to a semiclassical 
approximation to the Husimi function.

Since the zeroes of the Husimi function are the same of those of the
Bargmann function, the semiclassical distribution of the Husimi zeroes
can be obtained from (\ref{bargfunwkb}). 
However, besides the zeroes (\ref{zeroeswkb}) over the real axis, 
(\ref{bargfunwkb}) predicts spurious zeroes over the straight lines
that start in $z=z_{*}$ and have the directions $\theta=2\pi/3$ and
$\theta=4\pi/3$.   
Notwithstanding, considering the region of validity for applying 
(\ref{airyapp}) in (\ref{bargf}), we see that those zeroes are in a region 
where (\ref{bargfunwkb}) is not an approximation of the Bargmann
function (\ref{bargf}).



\newpage
\hspace{6.5cm}
CAPTIONS

%
%

{\bf FIG.1.}
Semiclassical geometry of the Wigner function in 1-D systems for a typical
smooth convex curve, $\cal E$, of constant energy. 
The {\bf full} {\it chord} near the Wigner caustic 
$\cal L$, corresponds to a {\it center} point outside it.
As the {\it center} moves through $\cal L$, a bifurcation occurs.
When the center lies on $\cal L$ ({\bf dashed} {\it chords}), a second 
{\it chord} is born. 
Notice that the phase space velocities at the tips of this new 
longest {\it chord} are parallel, cancelling the area (\ref{denom}), 
while for the shortest {\it chord} this does not happend.
Finally, when the {\it center} is inside $\cal L$, there are three 
{\it chords} ({\bf dotted}).
The other elements of the geometry are explained in the text.

%
%

{\bf FIG.2.}
Husimi function plots (on a logarithmic density scale) for an even 
eigenstate in the problem of a particle in a box with hard walls; 
the stress is in dark for the greatest value and in white for the 
lowest one.
({\bf a}) Calculated numerically from (\ref{husdef1}) with the function
$\langle \Omega_{\bf X}|\psi \rangle$ expressed in the position representation. 
({\bf b}) Asymptotic approximation for the error functions in (\ref{hubox})
given by (\ref{errofunp}).
({\bf c}) Smoothing of simplified Wigner function, (\ref{hupboxn}).
The center of the white spots represent the zeroes.
The horizontal white lines are the branches of the classical trajectory
$|p_n|=0.3$ between the limits of the box.
The points at the tips of this branches are the ``corners of the phase
space box'' cited in the text. 
The central rectangle encloses the intersection of the region 
$|\Im m(w)|< \Re e(w)$ with $\Re e(w) > 0$, for all the error functions in
(\ref{hubox}). The vertical lines in plot ({\bf c}) enclose the intersections 
of the same region for the error functions in (\ref{hupboxn}).

%
%

\begin{figure} 
\epsfxsize=16.0cm
\epsfbox[36 282 587 496]{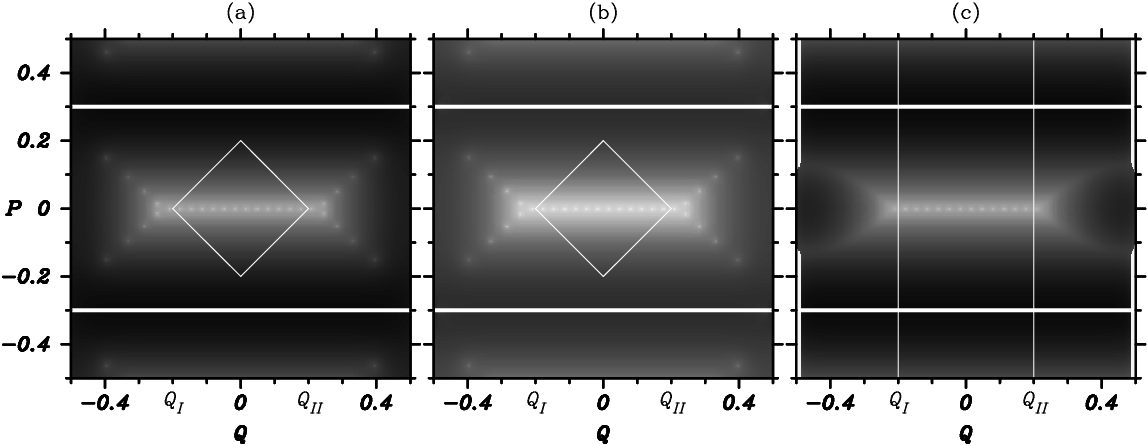}
\caption{}
\label{Fig2}
\end{figure}

\begin{figure} 
\epsfxsize=16.0cm
\epsfbox[82 208 481 554]{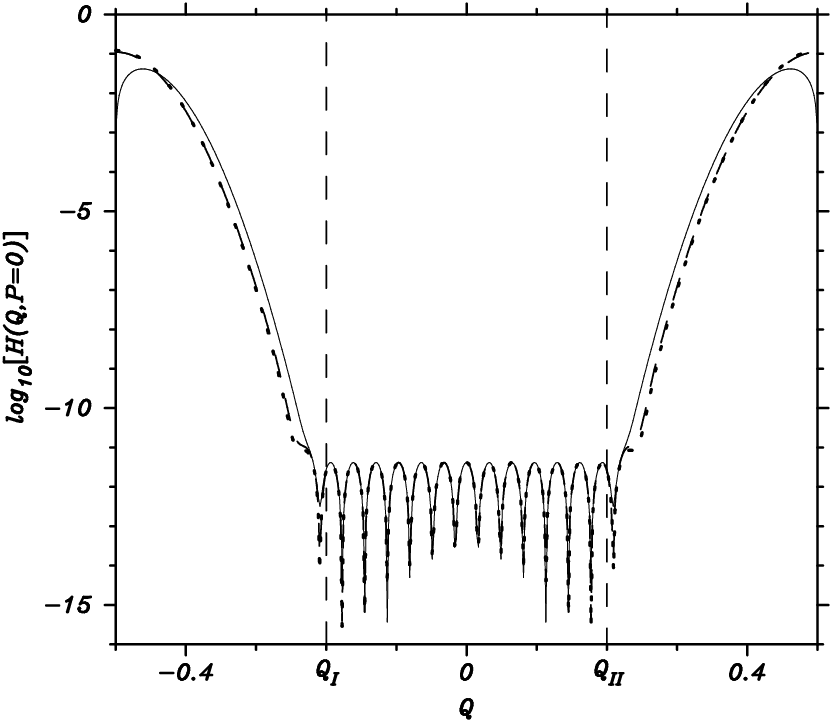}
\caption{}
\label{Fig3}
\end{figure}

\begin{figure} 
\epsfxsize=16.0cm
\epsfbox[47 214 538 552]{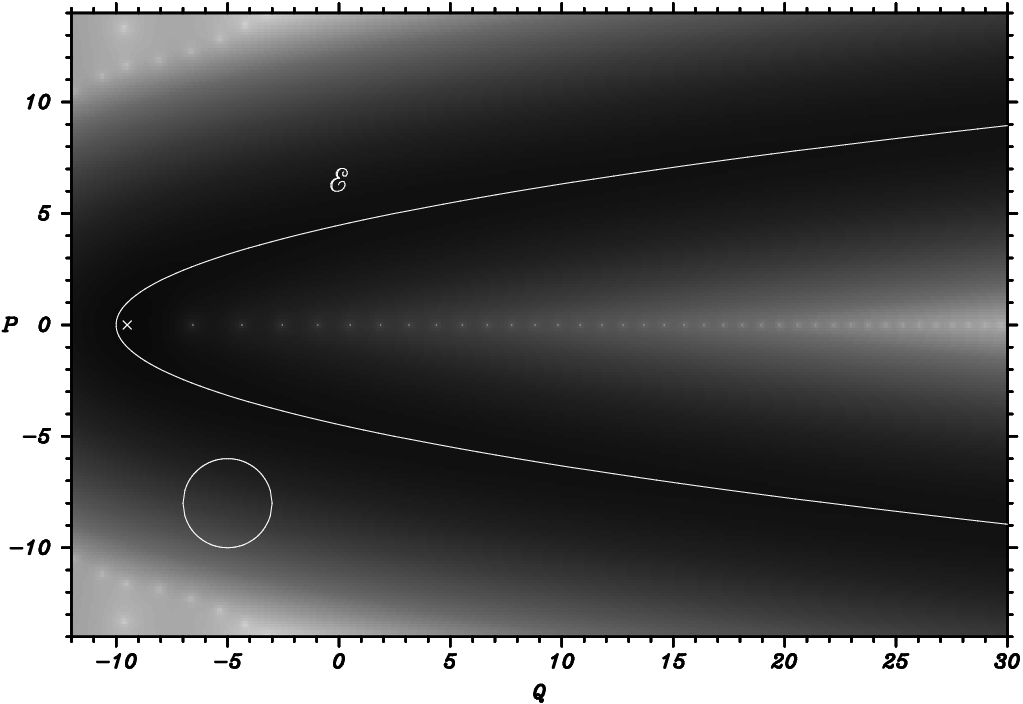}
\caption{}
\label{Fig4}
\end{figure}

\begin{figure} 
\epsfxsize=16.0cm
\epsfbox[32 170 538 592]{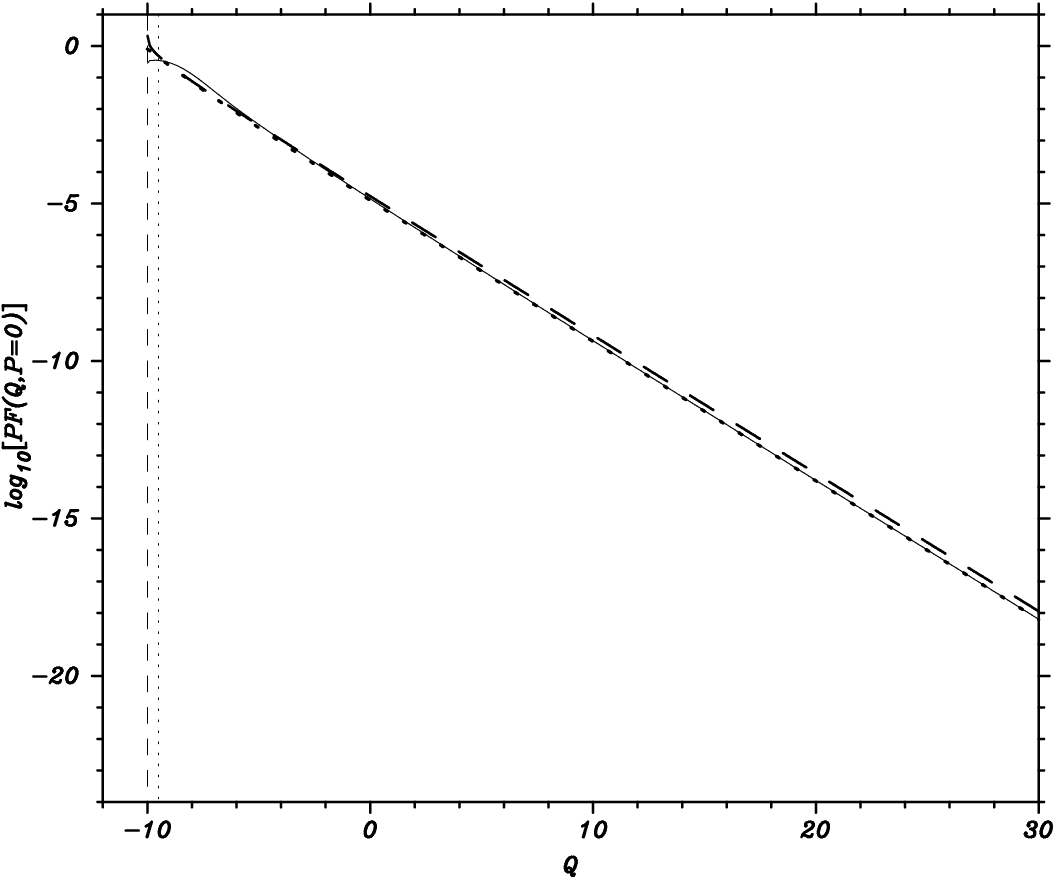}
\caption{}
\label{Fig5}
\end{figure}

\begin{figure} 
\epsfxsize=16.0cm
\epsfbox[32 170 538 592]{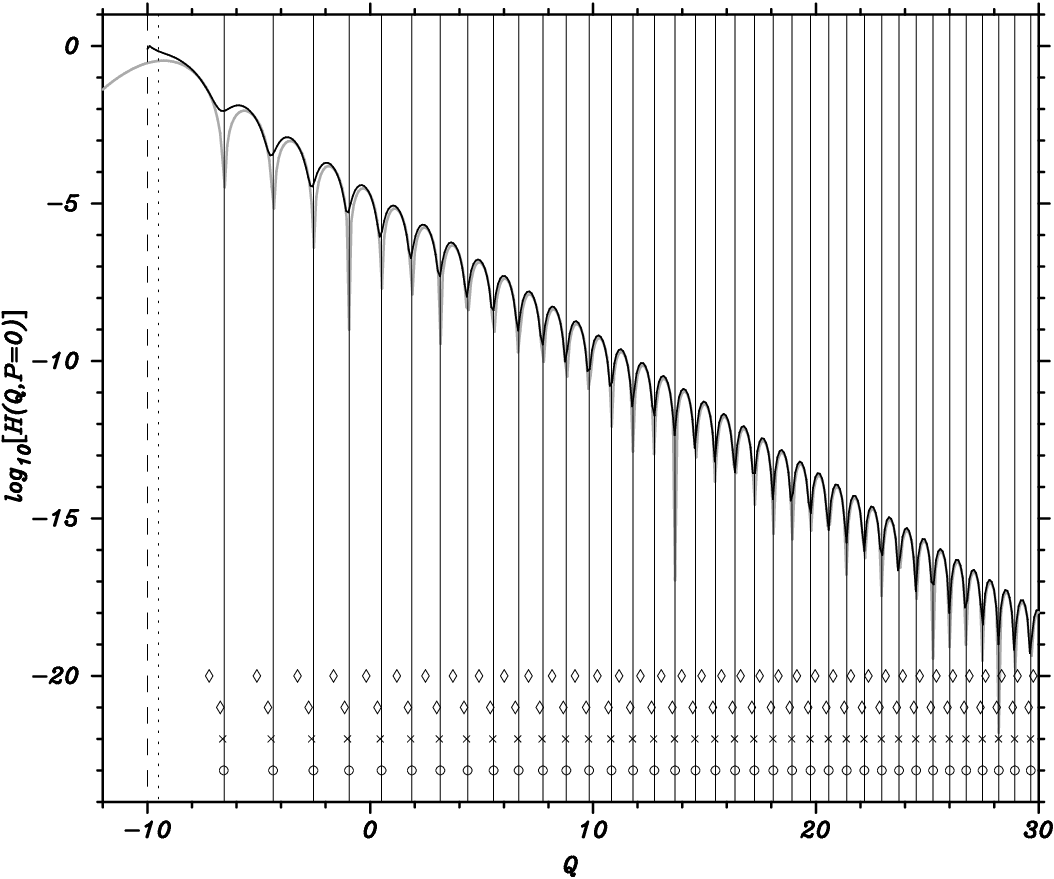}
\caption{}
\label{Fig6}
\end{figure}

\begin{figure} 
\epsfxsize=16.0cm
\epsfbox[82 247 481 516]{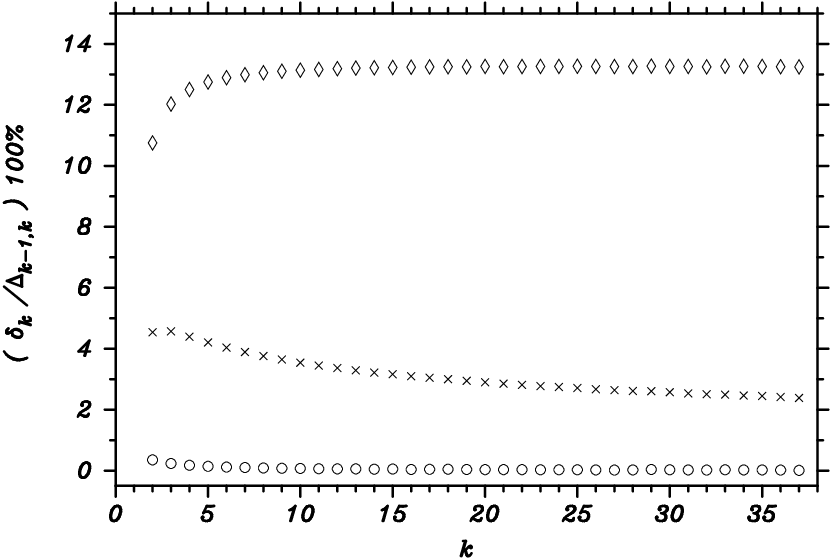}
\caption{}
\label{Fig7}
\end{figure}

{\bf FIG.3.}
The Husimi functions over the axis $P=0$ for the plots in FIG.\ref{Fig2}.
The {\bf dotted} curve corresponds to the plot ({\bf a}), the 
{\bf dashed} curve corresponds to the plot ({\bf b}) and the {\bf full}
line is for the plot ({\bf c}).  
The dashed vertical lines mark the limit of validity of
the approximation (\ref{hupebox}).

%
%

{\bf FIG.4.}
The Husimi function, on a logarithmic density scale, of an eigenstate for 
the particle under the action of a constant force.
The parabola $\cal E$, is the curve of constant energy.
The center of the white spots on the axis $P=0$ represent zeroes. 
The circle in the bottom left corner represent
the curve (\ref{elipse}) where the Gaussian smoothing (\ref{hugausmmoth}) is 
significative.  
The symbol ($\times$) in $P=0$ indicates the point $Q_{*}$. 
The values of the parameters used are: $E=10$, $F=1$, $\hbar=2$, 
$\beta=1$ and $m=1$.

%
%

{\bf FIG.5.}
Numerical camparison, along the $Q$-axis, of the prefactors of the cosine
in (\ref{SCLHUC}) ({\bf dotted} line) and in (\ref{SCLHUI}) 
({\bf dashed} line) with the integral (\ref{intfc}) ({\bf full} line).
Notice that  when $Q$ increases, the prefactor in (\ref{SCLHUI})
is grater than the integral (\ref{intfc}). 
The vertical dashed line is at position $q_r$ and the dotted 
one is at $Q_{*}$'s.

%
%

{\bf FIG.6.}
The logarithm of our approximation (\ref{SCLHUC}) 
({\bf black} curve) and of the Husimi function for the problem of the 
particle under the action of a constant force ({\bf gray} curve),
along the Q-axis inside the energy curve. 
The sharp inverted peaks indicate the position of the Husimi zeroes 
in the {\bf black} curve, and the minima of 
(\ref{SCLHUC}) in the {\bf gray} curve.  
The vertical full lines stress the position of the Husimi zeroes.
The vertical dashed line is at $q_r$'s position and the dotted 
one is at $Q_{*}$'s.
($\diamond$) indicates the $Q$-positions for the minima of our 
approximation (\ref{SCLHUI}) and ($\times$) for the minima
of (\ref{SCLHUC}).
The second line of ($\diamond$) are the minima of (\ref{SCLHUI})
shifted by the distance $(Q_{*}-qr)$.
($\circ$) are for the zeroes (\ref{zeroeswkb})
of the semiclassical Husimi function obtained by the WKB method in 
Bargmann representation.

%
%

{\bf FIG.7.}
Percentage relative error, $(\delta_k /\Delta_{k-1,k})100\%$ , between 
the positions of the Husimi zeroes (calculated numerically) and the 
positions given by the approximations to the zeroes.
$\delta_k$ is the distance, on the $Q$-axis, between  
the $k$-zero of the Husimi function and the position given by one of the
approximations.
$\Delta_{k-1,k}$ is the distance between the $k-1$ and the 
$k$-zero of the Husimi function. 
The zeroes are counting from left to right.
($\circ$) are for the case of the zeroes (\ref{zeroeswkb})
of the semiclassical Husimi function obtained by the WKB method in 
Bargmann representation, ($\times$) for the minima of our 
approximation (\ref{SCLHUC}) and the ($\diamond$) for those of 
our approximation (\ref{SCLHUI}) shifted by the distance $(Q_{*}-q_r)$.

%
%

{\bf FIG.8.}
Distribution of the zeroes of the Husimi functions of two energy
eigenstates inside the energy curve with $E \approx 30.8175$, 
for the problem of a particle subjet to an asymmetric anharmonic 
potencial (\ref{hamilanhar}).  
The symbol ($\times$) indicates the position of the Husimi zeros.
{\bf (a)}, for the eigenstate $n=30$ for a value of 
$\hbar \approx 0.508236$ and, {\bf (b)}, for the eigenstate $n=45$ 
for a value of $\hbar \approx 0.340691$.
$\cal E$ is the energy curve and $\cal L$ the Wigner caustic.
The {\bf dashed} line represents the {\it valley} of the Husimi zeroes.
The {\bf dotted} line represents the {\it valley} of local minima of
our approximation (\ref{SCLHUI}) and the {\bf full} line, the minima of 
(\ref{SCLHUC}). 
The circle at the upper right corner of each figure represents
the curve (\ref{elipse}) for the range of the Gaussian smoothing (\ref{hugausmmoth}).

%
%

{\bf FIG.9.}
Geometrical method for locating the position of the 
local minima, along the {\it valley} of approximations (\ref{SCLHUI})
{\bf (a)} and (\ref{SCLHUC}) {\bf (b)}.
The value of $\hbar$ corresponds to the quantum number $n=30$.
{\bf (a)}: {\bf dotted} lines indicate the level curves of the 
{\it center} action, while  the level curves of the 
{\it chord}'s length are given by the {\bf full} lines.
{\bf (b)}: {\bf dotted} lines indicate the level curves of the 
phase of the cosine in (\ref{SCLHUC}), while the level curves of
the argument of the exponential are given by the {\bf full} lines. 
{\bf (c)} and {\bf (d)}: the positions of the Husimi zeroes 
for the eigenstate $n=30$ ($\times$), and the approximate position 
of the local minima (($\diamond$) for (\ref{SCLHUI}), ($\circ$) for
(\ref{SCLHUC})).

%
%

{\bf FIG.10.}
Idem FIG.\ref{Fig9}, for a value of $\hbar$ corresponding to the quantum 
number $n=45$. 

%
%

\begin{figure} 
\epsfxsize=16.0cm
\epsfbox[31 172 588 660]{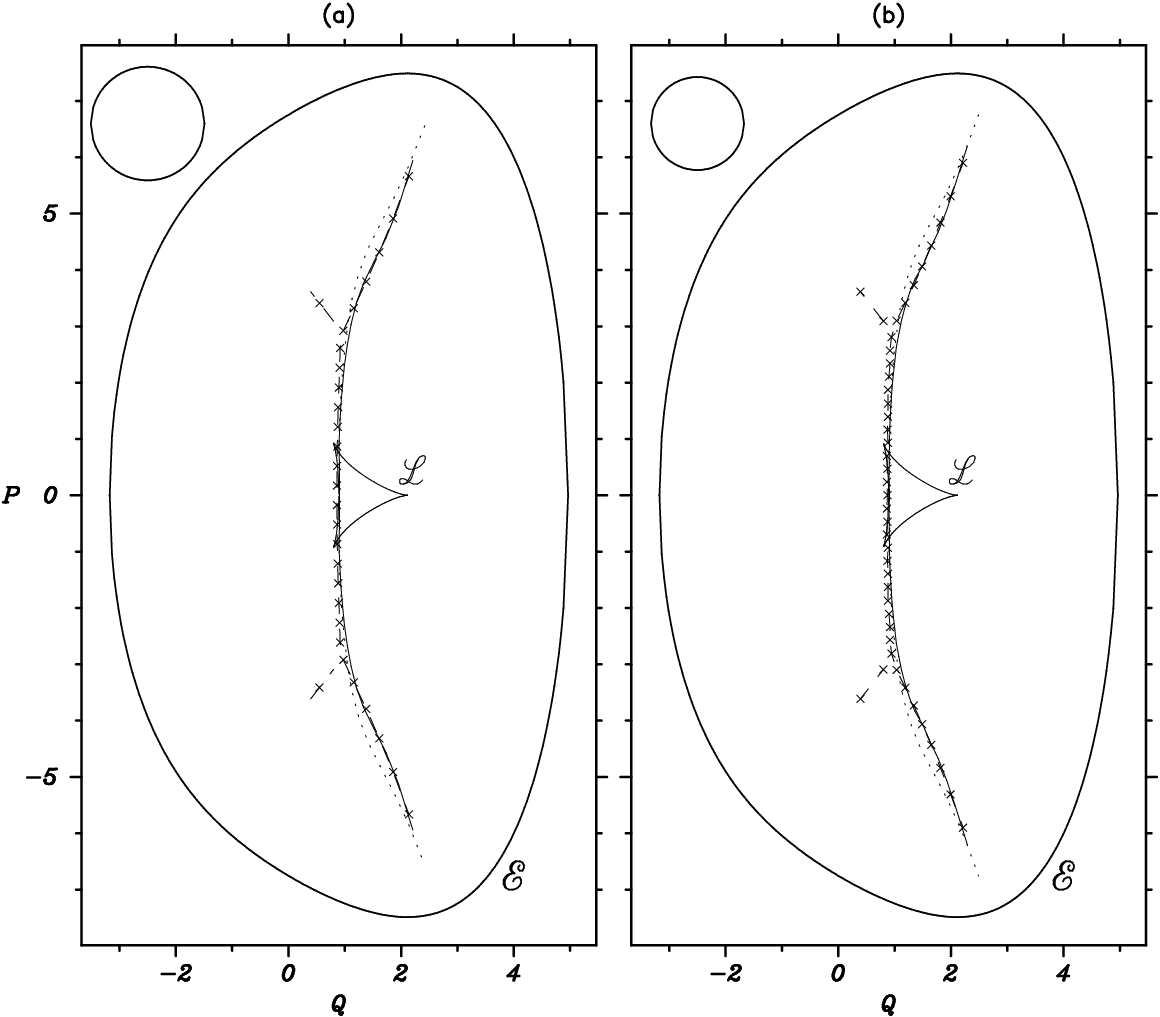}
\caption{}
\label{Fig8}
\end{figure}

\begin{figure} 
\epsfxsize=16.0cm
\epsfbox[42 148 554 630]{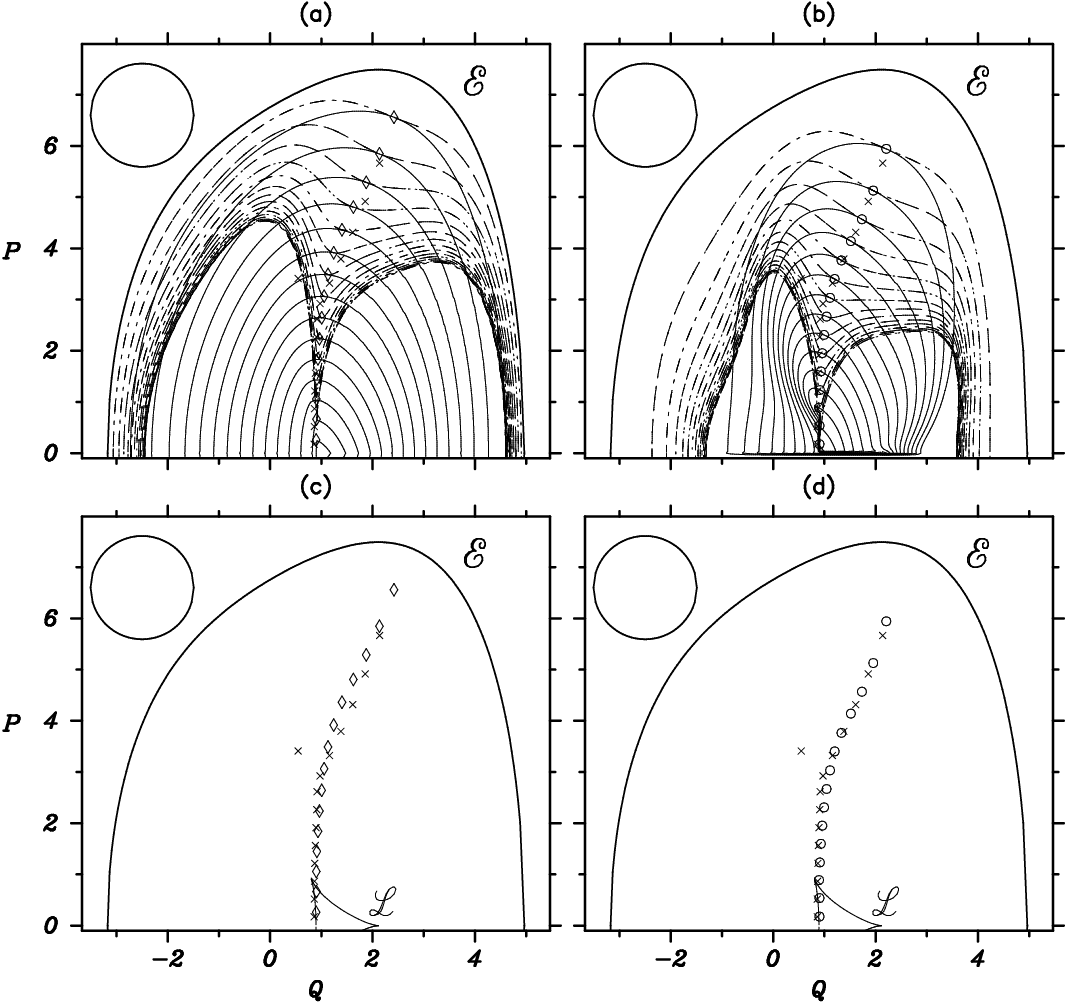}
\caption{}
\label{Fig9}
\end{figure}

\begin{figure} 
\epsfxsize=16.0cm
\epsfbox[42 148 554 630]{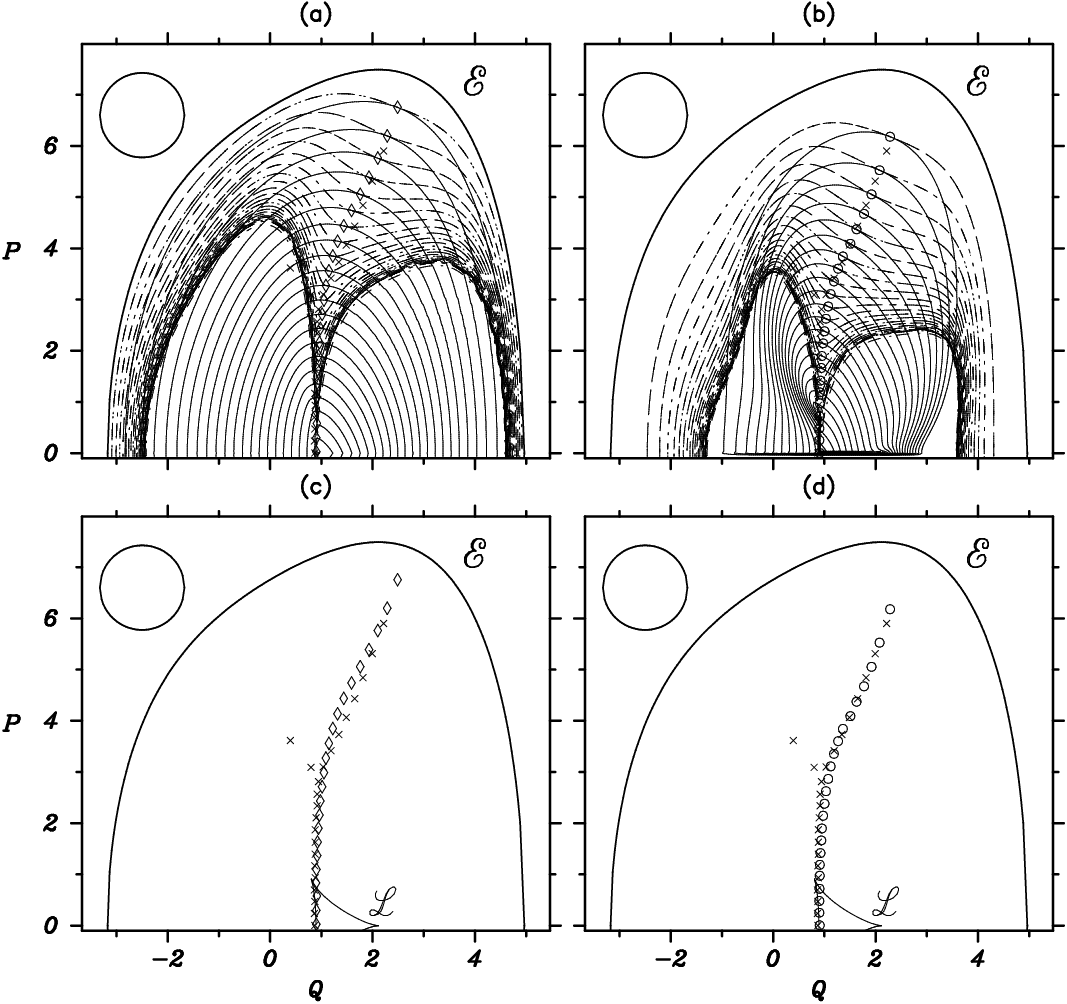}
\caption{}
\label{Fig10}
\end{figure}

{\bf FIG.11.}
The logarithm of our approximation (\ref{SCLHUC}) 
({\bf black} curve) and the Husimi function of the eigenstate
$n=30$ ({\bf gray} curve) along the {\it valley} of local 
minima ({\bf full} line in FIG.\ref{Fig8}).
The curves are projected onto the P-axis.
The vertical lines stress the position of the Husimi zeroes
over the P-axis.
The full vertical lines are for zeroes in the principal {\it valley}
of the Husimi function; the dashed vertical lines are for zeroes in
the shortest bifurcating{\it valley}.
The symbols ($\diamond$) and ($\circ$) correspond to the position
on the P-axis of the points in FIG.\ref{Fig9} {\bf (c)} and {\bf (d)}. 

%
%

\begin{figure} 
\epsfxsize=16.0cm
\epsfbox[26 172 537 592]{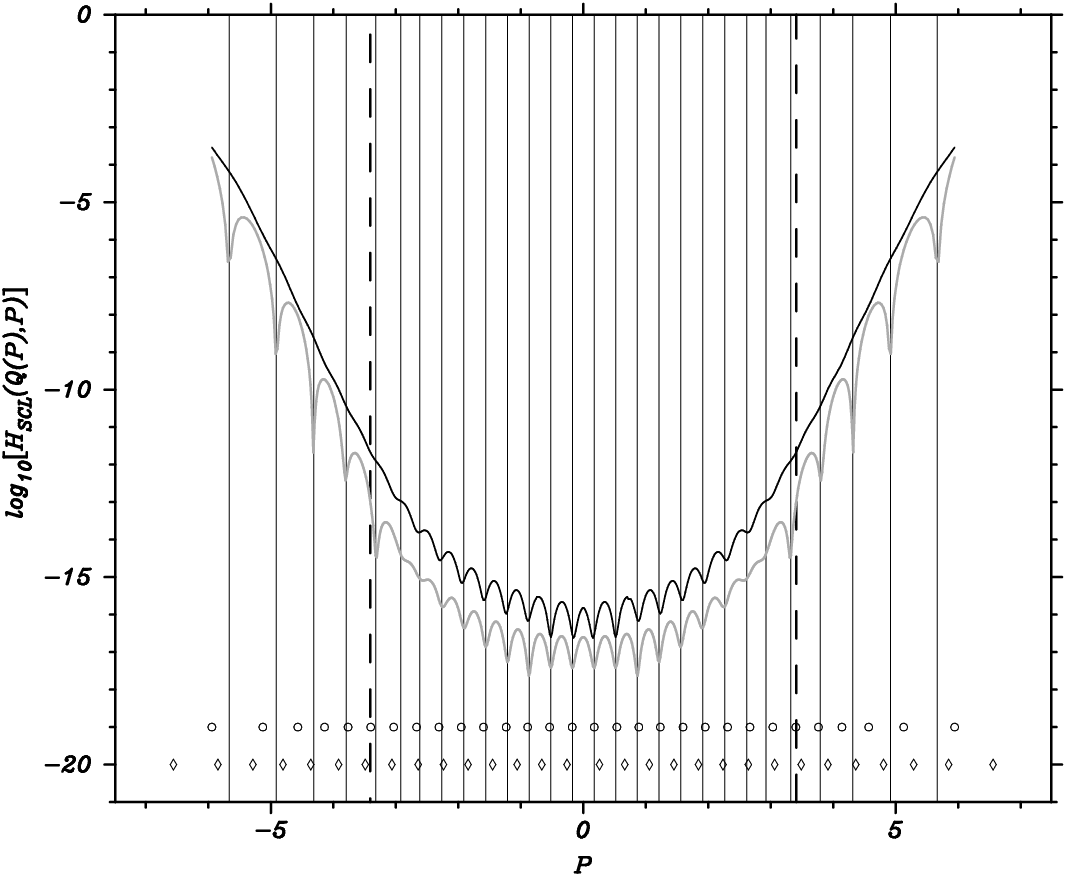}
\caption{}
\label{Fig11}
\end{figure}

\begin{figure} 
\epsfxsize=16.0cm
\epsfbox[26 172 537 592]{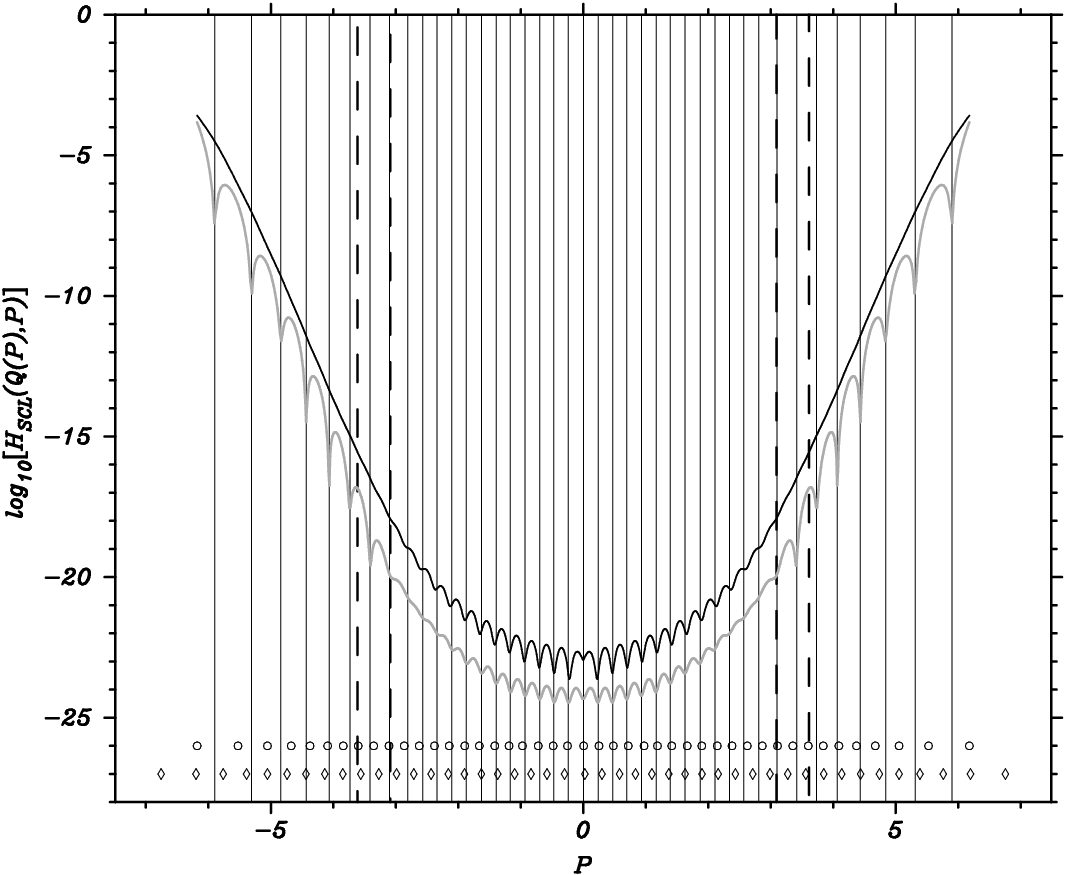}
\caption{}
\label{Fig12}
\end{figure}

{\bf FIG.12.}
Idem FIG.\ref{Fig11} for the eigenstate $n=45$. 


%
%
\newpage
%
%
\newpage

%
%
\newpage

%
%
\newpage

%
%
\newpage

%
%
\newpage

%
%
\newpage

%
%
\newpage

%
%
\newpage

%
%
\newpage

%
%
\newpage

%
%
\newpage



\end{document}